\documentclass[journal]{IEEEtran}

\ifCLASSINFOpdf
\else
\fi

\usepackage{hyperref}       
\usepackage{url}             
\usepackage{booktabs} 
\usepackage{amsfonts} 
\usepackage{nicefrac}        
\usepackage{microtype}       
\usepackage{bookmark}
\usepackage{array}
\usepackage{graphicx}
\usepackage{amsmath}
\usepackage{amsthm}
\usepackage{amsfonts} 
\usepackage{amssymb}
\usepackage{array}
\usepackage{algorithmic}
\usepackage{algorithm}
\usepackage{tabularx}
\usepackage{subfigure}
\usepackage{color}
\usepackage{cite}

\usepackage[T1]{fontenc} 
\usepackage{amsmath}
\usepackage[cmintegrals]{newtxmath}
\usepackage{bm} 

\usepackage{multirow}

\newtheorem{definition}{Definition}

\hypersetup{hidelinks}
\definecolor{airforceblue}{rgb}{0.36, 0.54, 0.66}
\definecolor{applegreen}{rgb}{0.55, 0.71, 0.0}
\definecolor{bittersweet}{rgb}{1.0, 0.44, 0.37}

\LetLtxMacro{\originaleqref}{\eqref}

\renewcommand{\eqref}{Eq.~\originaleqref}

\newcommand{\notice}{\color{black}}

\newcommand{\rv}{\color{black}}

\title{Safety-aware Semi-end-to-end Coordinated \\ Decision Model for Voltage Regulation \\ in Active Distribution Network}

\author{
  Linwei Sang,~\IEEEmembership{Graduate Student Member,~IEEE},
  Yinliang Xu$^*$,~\IEEEmembership{Senior Member,~IEEE},
  Huan Long,~\IEEEmembership{Member,~IEEE},
  and Wenchuan Wu,~\IEEEmembership{Fellow,~IEEE.}
  \thanks{This work was supported by National Nature Science Foundation of China, Grant No.~52277107, Shenzhen Science and Technology Program, Grant No.~JCYJ20210324130811031, Guangdong Basic and Applied Basic Research Foundation, Grant No.~2021A1515012450. Paper No.~TSG-00566-2022. (\textit{Corresponding Author: Yinliang Xu})}
  \thanks{Linwei~Sang, Yinliang~Xu are with Tsinghua-Berkeley Shenzhen Institute, Tsinghua Shenzhen International Graduate School, Tsinghua University, Shenzhen, China. {(E-mail: \url{sanglw21@mails.tsinghua.edu.cn}, \url{xu.yinliang@sz.tsinghua.edu.cn}.)}}
  \thanks{Huan~Long is with the School of Electrical Engineering, Southeast University, Nanjing, China. (E-mail:\url{hlong@seu.edu.cn})}
  \thanks{Wenchuan~Wu is with the Department of Electrical Engineering, Tsinghua University, Beijing, China. (E-mail: \url{wuwench@tsinghua.edu.cn})}
}

\begin{document}

\maketitle
\begin{abstract}
  Prediction plays a vital role in the active distribution network voltage regulation under the high penetration of photovoltaics. Current prediction models aim at minimizing individual prediction errors but overlook their collective impacts on downstream decision-making. Hence, this paper proposes a safety-aware semi-end-to-end coordinated decision model to bridge the gap from the downstream voltage regulation to the upstream multiple prediction models in a coordinated differential way. The semi-end-to-end model maps the input features to the optimal var decisions via prediction, decision-making, and decision-evaluating layers. It leverages the neural network and the second-order cone program (SOCP) to formulate the stochastic PV/load predictions and the var decision-making/evaluating separately. Then the var decision quality is evaluated via the weighted sum of the power loss for economy and the voltage violation penalty for safety, denoted by \textit{regulation loss}. Based on the \textit{regulation loss} and prediction errors, this paper proposes the hybrid loss and hybrid stochastic gradient descent algorithm to back-propagate the gradients of the hybrid loss with respect to multiple predictions for enhancing decision quality. Case studies verify the effectiveness of the proposed model with lower power loss for economy and lower voltage violation rate for safety awareness.
\end{abstract}

\begin{IEEEkeywords}
  Semi-end-to-end learning, voltage regulation, neural network, active distribution network, prediction model, safety awareness.
\end{IEEEkeywords}
  
\IEEEpeerreviewmaketitle

\section{Introduction}\label{sec: intro}

\IEEEPARstart{T}{o} alleviate the carbon emission, the penetration of solar photovoltaics (PVs) keeps proliferating in the distribution networks during recent years \cite{Victoria2021}. However, the intermittence/stochasticity of PV and the high R/X ratios of the distribution network require active voltage regulation to enhance the system operation safety. The regulation decision-making relies on the multiple PV/load predictions to solve the formulated voltage regulation optimization problem, which follows the predict-and-optimize framework \cite{Elmachtoub2021}. Under this framework, various PV prediction methods for prediction accuracy and voltage regulation models for decision quality have been proposed in recent years.

Energy prediction is a key element for accommodating the increasing renewable energy penetration in the power system. State-of-the-art PV prediction methods leverage comprehensive data sources \cite{Kim2019, Catalina2020, Chang2020} and burgeoning machine learning \cite{Ahmed2020, Liu2018, Simeunovic2022, Cheng2022-1, Cheng2022-2, Yao2022, Wang2019} techniques to improve PV prediction accuracy. Ref. \cite{Kim2019} analyzes the significance of input features for the PV predictive models. Further, the combination of numerical weather predictions and satellite data can facilitate the SVR-based PV prediction models effectively in Ref. \cite{Catalina2020}. Ref. \cite{Ahmed2020} reviews the PV prediction research frontier from the perspectives of the data preprocessing, model construction, and model evaluation. A novel two-stage model is proposed in Ref. \cite{Liu2018} for predicting power and quantifying the uncertainty of prediction models. Spatio-temporal graph neural networks are proposed in Ref. \cite{Simeunovic2022} for multi-site PV forecasting, capturing the intuition of the dependencies of PV production data for multiple PVs' prediction accuracy. {\rv An end-to-end short-term PV forecasting model takes the satellite images as inputs to learn the cloud motion for multistep ahead prediction \cite{Cheng2022-1}, and further, a graphical learning framework is proposed for intra-hour PV power prediction based on the images in Ref. \cite{Cheng2022-2}. An advanced U-net and an encoder-decoder architecture are integrated into deep learning to enhance prediction accuracy \cite{Yao2022}. Key weather information is explored in Ref. \cite{Wang2019} to improve the prediction accuracy. Rather than the direct energy prediction, Ref. \cite{Toube2022} focuses on the system imbalance prediction based on the interpretable recurrent neural networks.} Current research focuses on prediction accuracy by minimizing the prediction errors but overlooks the multiple prediction collective impact on the decision quality of downstream decision-making. This paper focuses on exploiting the full potential of coordinated prediction models for enhancing decision quality.

Multiple PV and load predictions instruct the downstream decision-making for voltage regulation \cite{Zhang2020}. A novel hierarchically-coordinated VCC method is proposed to dispatch the reactive inverter power for minimizing power loss by the scenario-based stochastic formulation of PV and load uncertainty in Ref. \cite{Zhang2020}. A robust optimization model for considering the worst PV scenario is formulated and solved by the proposed distributed adaptive robust voltage/var control method in Ref. \cite{Li2020}. A two-stage chance-constrained formulation of distributed generation (PV, wind power) and load uncertainties is proposed in Ref. \cite{Nazir2019} for volt/var regulation. Ref. \cite{Liu2021} proposes an online multiagent reinforcement and decentralized control framework, which embeds the uncertainty in the data-driven control model. {\rv A distributed control strategy for PV inverters is proposed in Ref. \cite{Long2022} to promote the voltage regulation capacity. Ref. \cite{Zhang2020-se} coordinates the reactive power between the PV and battery storage using an event-triggered approach. An incentive-based fairness approach is proposed to mitigate fast voltage deviation with the fairness guarantee in Ref. \cite{Sun2022}.} The above optimization-based decision-making models formulates the uncertainties of PV predictions by stochastic \cite{Zhang2020,Agalgaonkar2015}, robust \cite{Li2020}, chance-constrained \cite{Nazir2019}, and data-driven \cite{Li2022, Liu2021, Liu2022, Chen2020} methods, which seldom consider the reverse impact of decision-making on the multiple prediction models for learning to enhance decision quality.

The above researches indicate the gap between downstream decision-making and upstream predicting. Conventional voltage regulation decision-making under PV penetration follows a two-stage pipeline. In the first stage, the renewable energy prediction by learning models aims at minimizing prediction errors \cite{Kim2019, Catalina2020, Chang2020, Ahmed2020, Liu2018, Simeunovic2022}, and in the second stage, based on the predicted values, voltage regulation decisions are decided by optimization models \cite{Zhang2020, Li2020, Agalgaonkar2015, Nazir2019, Liu2021, Chen2020}. Under the above pipeline, the connection between the two stages is unidirectional. Prediction focuses on the {\rv prediction accuracy} by minimizing the prediction errors between the predicted and ground truth values, while decision-making focuses on the {\rv decision quality} by optimizing the decisions based on the predicted values. {\rv So we propose to train multiple prediction models not solely for increasing prediction accuracy but for enhancing decision quality. This formulation may seem counter-intuitive, given that ``perfect'' prediction models would lead to optimal decision-making. However, the reality that all models do inherit errors illustrates that we should indeed emphasize a final decision-quality objective to determine the proper error trade-offs within a machine learning setting \cite{Donti2017, Lu2020, Sang2022}.} The objectives of {\rv the prediction and decision-making stages} are not the same, and under the same {\rv prediction accuracy}, the {\rv decision quality} can be further improved. 

To address the above inconsistency, we propose the safety-aware semi-end-to-end coordinated decision model for voltage regulation, which considers the {\rv decision quality} by integrating the downstream regulation model in training the upstream multiple PV/load prediction models via a coordinated differential way. The semi-end-to-end decision model leverages the neural network (NN) models (intractable) to formulate the stochastic PV/load power and the second-order convex program (SOCP) models (tractable) to formulate the optimal var decision-making. Its main framework includes three layers, mapping from the input features to multiple PV/load predictions by the NN-driven prediction layer, from predictions to var decisions by the SOCP-driven decision-making layer, and from the {\rv var} decisions to the decision quality evaluating by the SOCP-driven decision-evaluating layer. It evaluates the var decision quality under predictions by the weighted sum of the power loss for economy and the voltage violation penalty for safety awareness, denoted by \textit{regulation loss}. The gradients of the \textit{regulation loss} with respect to the predicted PV/load are derived from the SOCP-driven decision-making and evaluation through SOCP derivative solution map, including skew-symmetric mapping, homogeneous self-dual embedding, and solution construction. Combined with prediction errors, we propose the safety-aware semi-end-to-end hybrid stochastic gradient descent (HSGD) learning algorithm to 
back-propagate the hybrid gradients to multiple prediction models for reducing \textit{regulation loss} and enhancing the {\rv decision quality}. So the principal contributions of this paper can be summarized threefold:

{\rv
1) To the authors' best knowledge, this paper, for the first time, takes advantage of the downstream voltage regulation model to train the multiple prediction models in a coordinated differential way for enhancing decision quality. Compared to conventional prediction approaches \cite{Kim2019, Catalina2020, Chang2020, Ahmed2020, Liu2018, Simeunovic2022}, the proposed semi-end-to-end model utilizes the reverse impact of downstream decision-making on the multiple upstream predictions to improve the decision quality under the multiple coordinated predictions.

2) \textit{Regulation loss} is proposed to evaluate the decision quality of voltage regulation, including the economic part for minimizing power loss and the safety-aware part for guaranteeing safe operation. The gradients of \textit{regulation loss} to multiple predictions are derived by SOCP derivative solution map and back-propagated to the multiple prediction models for improving decision quality.

3) Based on the \textit{regulation loss} and prediction errors, this paper proposes the hybrid loss and the corresponding safety-aware semi-end-to-end HSGD learning algorithm to train the multiple prediction models. The case study verifies the economy and safety awareness of the proposed decision model with lower power loss and lower voltage violation rate.
}

{\rv 
To summarize, the main novelty of this paper is to utilize the downstream SOCP-driven voltage regulation model to train the upstream multiple NN-driven renewables and load prediction models for enhancing not only prediction accuracy but the decision quality. It bridges the gap between SOCP-driven decision-making and NN-driven prediction for the first time.

It should be noted that the degrading of decision-making is due to the uncertainties of prediction. The uncertainties can come from two aspects: i) energy prediction \cite{Zhang2020, Li2020, Agalgaonkar2015, Nazir2019}: the uncertain prediction of renewable generation and load demand due to their intrinsic stochasticity and volatility; ii) system estimation \cite{Massignan2022,Zhang2020-se}: the uncertain estimation of system parameters (e.g., line impedance) due to bad data and measuring noise. This work focuses on improving the decision quality under the energy prediction uncertainty by assuming the state estimation is accurate, which is consistent with other state-of-the-art works \cite{Zhang2020, Li2020, Agalgaonkar2015, Nazir2019}.}

The rest of this paper is organized as follows. Section \ref{sec: prob} presents the conventional predict-and-optimize voltage regulation {\rv preliminaries}. Section \ref{sec: method} proposes the safety-aware semi-end-to-end coordinated decision model for voltage regulation. Section \ref{sec: case} performs the case study to verify the effectiveness of the proposed decision model. Section \ref{sec: conclusion} concludes the paper.


{\rv
\section{Conventional Predict-and-optimize Voltage Regulation Preliminaries} \label{sec: prob}
}

The section introduces the conventional predict-and-optimize voltage regulation {\rv preliminaries}, composed of two stages: 1) PV/load prediction based on the feature context for reducing prediction errors; 2) the voltage regulation optimization based on the predicted PV outputs for minimizing the power loss.

\subsection{Upstream PV/load Prediction Models} 

Multiple PV systems are usually integrated into the different buses of the distribution network. Due to PV fluctuation and intermittence, voltage regulation is essential for constraining the voltage magnitude and reducing power loss. Day-ahead voltage regulation strategy relies on the PV/load predictions in the different buses. So we formulate the PV/load prediction models in \eqref{eq: PV prediction model}.
\begin{eqnarray}
  \begin{aligned}
    \hat{p}^{pred}_i = f^{pred}_{\theta_i}(x_i), \quad i \in \mathcal{N}_{pred}.
  \end{aligned}
  \label{eq: PV prediction model}
\end{eqnarray}
where $\hat{p}^{pred}_i$ is the day-ahead predicted power output of prediction model $i$; $x_i$ is the feature context of model $i$; $f^{pred}_{\theta_i}(\cdot)$ is the prediction model $i$ parameterized by $\theta_i$; $\mathcal{N}_{pred}$ is the set of prediction models, located in different buses. It should be noted that we can replace $(\cdot)^{pred}$ with $(\cdot)^{PV}$ for PV prediction and $(\cdot)^{load}$ for load prediction.

Then we utilize the neural network (NN) to formulate the mapping relationship from the prediction feature context $x_i$ to the predicted output $\hat{p}^{pred}_i$ by $f^{pred}_{\theta_i}(\cdot)$ and then train it to learn the optimal parameter $\theta_i$ by stochastic gradient descent for minimizing the mean-square-error (MSE) of predicted PV/load, as illustrated in \eqref{eq: mse training}. {\rv The detailed stochastic gradient descent training and NN structure are attached in the appendix \ref{subsec: nn struct}.}

\begin{subequations}
  \begin{align}
    \min_{\theta_i} \quad & \frac{1}{|\mathcal{N}|}\sum_{n \in \mathcal{N} }\sum_{t \in \mathcal{T}} (\hat{p}^{pred}_{i,t} - p^{pred}_{i,t})^2 \\
    \text{s.t.} \quad & \hat{p}^{pred}_i = f^{pred}_{\theta_i}(x_i).
  \end{align}
  \label{eq: mse training}
\end{subequations}

Each prediction model focuses on reducing each individual prediction error for prediction accuracy. After training, the multiple PV and load predictions are collectively delivered to the downstream parameterized SOCP-driven voltage regulation model for decision-making.

\subsection{Downstream Parameterized SOCP-driven Voltage Regulation Model}

The parameterized SOCP-driven voltage regulation model takes the multiple PV predictions as the input parameters $\hat{p}^{PV}, \hat{p}^{D}, \hat{q}^{D} \in \mathcal{X}$ to make the optimal var regulation decisions $\hat{q}^{reg} \in \mathcal{Y}$ in \eqref{eq: voltage regulation opt}, detailed in Ref.\cite{Li2020}. It should be noted that the var regulation capacity can be from the PV inverters or SVC, which is not the focus of this paper.

\begin{subequations}
  \small
  \allowdisplaybreaks
  \begin{align}
    \min_{q^{reg}} \quad & \qquad \sum_{t\in \mathcal{T}} \sum_{ij \in \mathcal{E}} l_{ij,t} r_{ij} \label{socp a}\\
    \text{s.t.} \quad &\sum_{jk \in \mathcal{E}} P_{jk,t} - \sum_{ij \in \mathcal{E}} (P_{ij, t} - r_{ij}l_{ij,t}) = \hat{p}^{PV}_{j,t} - \hat{p}^{D}_{j,t} \label{socp b}\\
    & \sum_{jk \in \mathcal{E}} Q_{jk,t} - \sum_{ij \in \mathcal{E}} (Q_{ij, t} - x_{ij}l_{ij,t}) = \hat{q}^{reg}_{j,t} - \hat{q}^{D}_{j,t} \label{socp c}\\
    & v_{j,t} = v_{i,t} + (r^2_{ij}+x^2_{ij})l_{ij,t} - 2 (r_{ij}P_{ij,t} + x_{ij}Q_{ij,t}) \forall ij \in \mathcal{E} \label{socp d}\\
    & || 2P_{ij,t} \quad 2Q_{ij,t} \quad l_{ij,t}-v_{i,t} || \leq l_{ij,t} + v_{i,t} \label{socp e}\\
    & (V^{min}_{i,t})^2 \leq v_{i,t} \leq (V^{max}_{i,t})^2, \forall i \in \mathcal{N}_B \label{socp f}\\
    & (I^{min}_{ij,t})^2 \leq l_{ij,t} \leq (I^{max}_{ij,t})^2, \forall ij \in \mathcal{E} \label{socp g} \\
    & q^{reg, min}_{i,t} \leq q^{reg}_{i,t} \leq q^{reg, max}_{i,t}, \forall i \in \mathcal{N}_{reg}. \label{socp h}
  \end{align}
  \label{eq: voltage regulation opt}
\end{subequations}
where $v_{i,t}$, $l_{ij,t}$ are the square of voltage magnitude and current; $P_{ij,t}$ and $Q_{ij,t}$ are the active and reactive power of branches; $\mathcal{T}$ and $\mathcal{E}$ are the set of operation periods and network branches; $r_{ij}$ and $x_{ij}$ are the resistance and reactance of branch $ij$. {\rv The detailed presentation of \eqref{eq: voltage regulation opt} is attached in appendix \ref{subsec: vcc}.}

Unlike explicit functions mapping from the input space $\mathcal{X}$ to the output space $\mathcal{Y}$ directly, the parameterized voltage regulation model maps the PV predictions to the regulation decisions implicitly via the SOCP of \eqref{eq: voltage regulation opt}. {\rv It should be noted that the varying parameters in this paper are only uncertain PV generation and load at buses and time periods across instances; in contrast, the generator costs, distribution network topology/parameters, and other parameters are assumed to be fixed across instances.}

\subsection{Gap Between the Multiple Prediction and Optimization}

The conventional predict-and-optimize framework indicates two gaps: 1) upstream predictions flow to the downstream optimization unidirectionally, 2) and PV/load prediction models are trained based on local historical information for prediction accuracy separately.

This paper considers the collective prediction error impact on the downstream voltage regulation model and trains the multiple prediction models for enhancing decision quality. In the process, multiple PV and load predictions are integrated into the SOCP-driven voltage regulation model, so the utilization of the regulation model enables the coupling of multiple prediction models to facilitate the system operation. So the safety-aware semi-end-to-end coordinated decision model is proposed to bridge the above gaps in the following section.

\section{Methodology}\label{sec: method}

This section introduces the semi-end-to-end coordinated decision model for voltage regulation to bridge the gap from voltage regulation to multiple PV and load predictions via gradients of the decisions with respect to the predictions. It first presents the semi-end-to-end learning framework, then proposes the semi-end-to-end coordinated decision model, evaluates the decision quality, and presents the implicit solution map of SOCP derivative for training the prediction models in a coordinated differential way.

\subsection{Semi-end-to-end Learning Framework}

End-to-end learning usually refers to learning the optimal decisions directly from raw input feature context by machine learning methods \cite{Muller2005}. The direct application of end-to-end learning in voltage regulation takes the conventional predict-then-optimized framework as a black-box model. It utilizes complex deep learning to capture the mapping from $x_i$ to $\hat{q}^{reg}$. Nevertheless, it ignores the fact that the mapping from $x_i$ to $\hat{p}^{PV}, \hat{p}^{D}, \hat{q}^{D}$ is intractable due to the solar irradiance and load behaviors. In contrast, the mapping from $\hat{p}^{PV}, \hat{p}^{D}, \hat{q}^{D}$ to $\hat{q}^{reg}$ can be formulated by the tractable SOCP {\rv problem}. 

Unlike end-to-end learning, the semi-end-to-end coordinated decision model for voltage regulation is proposed by considering the intractable multiple PV/load {\rv prediction} and the tractable voltage regulation decision models. Ref. \cite{Kotary2021} surveys the emerging end-to-end constrained optimization learning advancements {\rv with} focus on integrating the optimization methods with the machine learning architecture. Then we define the semi-end-to-end decision model in a general form as follows:

\begin{definition}[Semi-end-to-end Decision Model]\label{def: semie2e}
  A semi-end-to-end decision model $\hat{y}_\theta: \mathcal{X} \rightarrow \mathcal{Y}$ maps the input parameters to the solution of optimization problem and accesses the objective of problem iteratively, which adds the domain knowledge of optimization into $\hat{y}_\theta (x)$, illustrated by \eqref{eq: semie2e def}.
\end{definition}

\begin{equation}
  \hat{y}^0_\theta \rightarrow  \hat{y}^1_\theta \rightarrow \cdot \cdot \cdot \rightarrow \hat{y}^K_\theta := \hat{y}_\theta (x)
  \label{eq: semie2e def}
\end{equation}
where $\hat{y}^{(\cdot)}_\theta$ is the intermediate variables and $\hat{y}_\theta$ is final decision variables, where the mappings in \eqref{eq: semie2e def} can be formulated by machine learning or optimization techniques.

After formulating the stacked mapping relationship, the objective of the loss function is formulated to learn the parameters $\theta$ of $\hat{y}_\theta$. According to the section III of Ref. \cite{Amos2022}, learning the $\theta$ of \eqref{eq: semie2e def} can be categorized into 1) matching a ground-truth solution, where regression-based loss function minimize the distance between the prediction decision $\hat{y}_{\theta}(x)$ and the ground truth decision $y^{*}(x)$ in \eqref{eq: obj a}, and 2) minimizing the objective of \eqref{eq: obj b}, where the objective-based loss function learns the $y^{*}(x)$ with the local information of the objective $f$ and without access to the $y^*$. 

\begin{subequations}
  \begin{align}
    \arg \min_{\theta} \quad &\mathcal{L}_{reg}(\hat{y}_{\theta}) = \mathbb{E}_{x \sim \mathcal{P}(x)} ||\hat{y}_{\theta}(x) - y^*_{\theta}(x)|| \label{eq: obj a}\\
    \arg \min_{\theta} \quad &\mathcal{L}_{obj}(\hat{y}_{\theta}) = \mathbb{E}_{x \sim \mathcal{P}(x)} f(\hat{y}_{\theta}(x); x).  \label{eq: obj b}
  \end{align}
  \label{eq: e2e obj}
\end{subequations}

Inspired by the above, this paper utilizes an objective-based loss function design to evaluate the decision quality of voltage regulation for the proposed semi-end-to-end coordinated decision model. So the keys of the semi-end-to-end learning model lie in two aspects: 1) \textit{forward propagation:} design proper loss function in \eqref{eq: e2e obj}, and integrate the optimization models into learning process in \eqref{eq: semie2e def}; 2) \textit{backward propagation:} derive the gradients from the above loss function/ optimization models, and back-propagate the gradients for updating $\theta$ of $\hat{y}_{\theta}$. 

The proposed semi-end-to-end model requires differentiating through vector-valued functions to vector arguments. So we clarify our notions related to differentiating for clear delivery. For the function $f(x, y, z)$ where $y$, $z$ are the functions of $x$, then we have: 

\begin{eqnarray}
  \begin{aligned}
    \mathsf{D}_{x} f &=  \frac{\partial f}{\partial x} \frac{d x}{d x} + \frac{\partial f}{\partial y}\frac{d y}{d x} + \frac{\partial f}{\partial z}\frac{d z}{d x} \\
    &=  \mathsf{D}_x f \mathsf{D}_x x + \mathsf{D}_y f \mathsf{D}_x y + \mathsf{D}_z f \mathsf{D}_x z .
  \end{aligned}
  \label{eq: diff im}
\end{eqnarray}

We take $\mathsf{D}_{(\cdot)}$ as the derivative operation with respect to subscript variable $(\cdot)$.

Then the chain rule for $h(x) = f(g(x))$ is presented as:
\begin{equation}
  \mathsf{D}_{x}h(x) = \mathsf{D}_{g}f(g(x)) \mathsf{D}_{x}g(x).
  \label{eq: diff chain}
\end{equation}
The subscript of $(\cdot)$ can be dropped when the input variable is declared explicitly, so we can reach that $\mathsf{D}h(x)$ is the same as $\mathsf{D}_{x}h(x)$.

\subsection{Safety-aware Semi-end-to-end Coordinated Decision Model for Voltage Regulation}

\begin{figure}[ht]
  \centering
  \rv
  \includegraphics[scale=1.15]{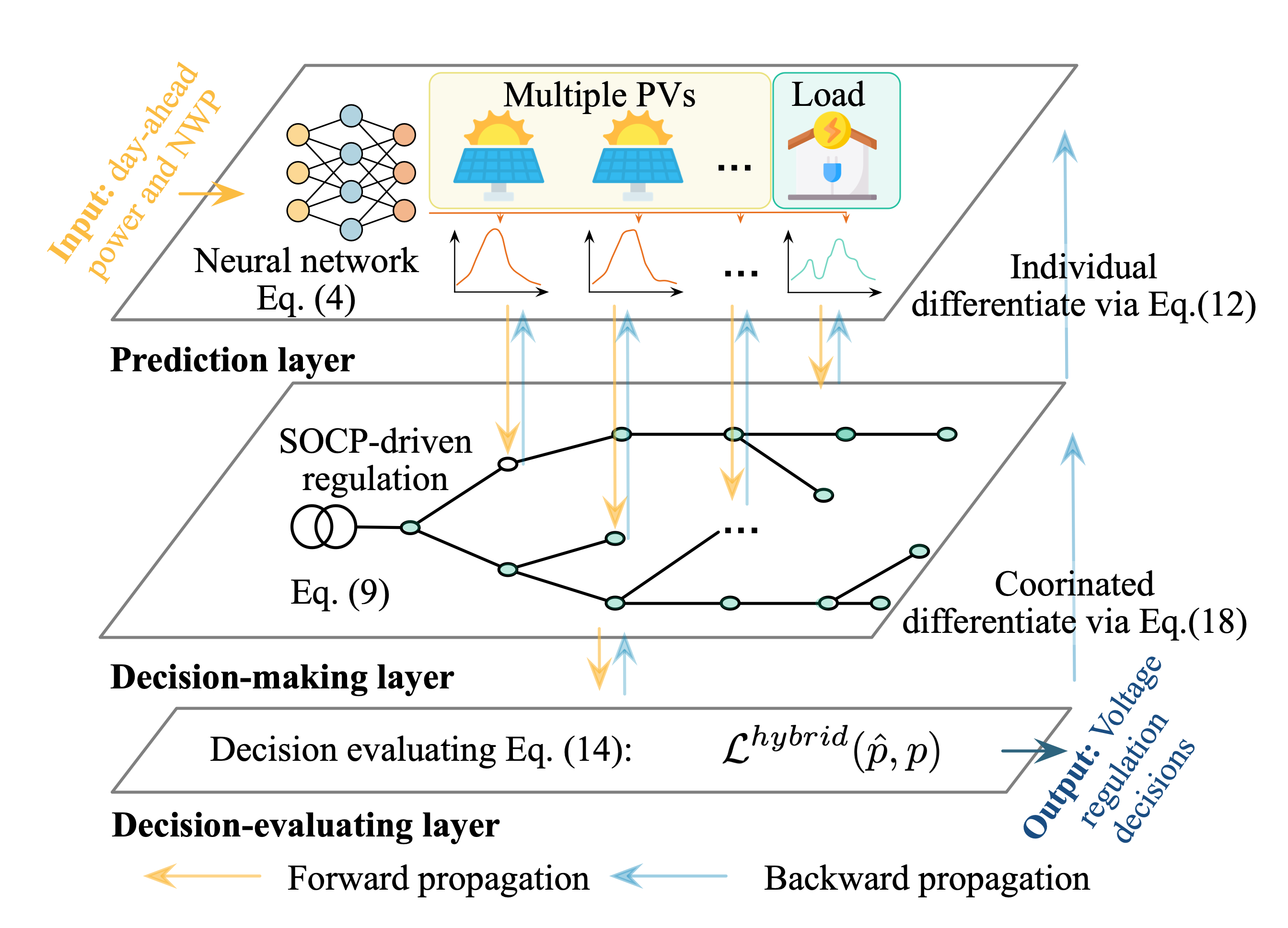}
  \caption{Safety-aware semi-end-to-end coordinated decision model for voltage regulation.}
  \label{fig: se2e framework}
\end{figure}

Inspired by the above semi-end-to-end idea of \eqref{eq: obj b}, we propose a safety-aware semi-end-to-end coordinated decision model for voltage regulation via integrating the downstream regulation model into training the upstream multiple prediction models in a coordinated differential way, which is composed of three layers with coordinated differential forward / backward propagation procedures, as shown in Fig. \ref{fig: se2e framework}. {\rv It takes the day-ahead power (PV/load) and numerical weather prediction (NWP) information as input and provides the voltage regulation decisions as output.}

In the coordinated \textit{forward} propagation, the prediction (first) layer maps the input feature contexts to PV/load predictions by NN (\ref{eq: nn struct}); the decision-making (second) layer maps the multiple predictions to var regulation decisions by the SOCP-driven voltage regulation model (\ref{eq: voltage regulation opt}); the decision-evaluating (third) layer maps the var decisions to the hybrid loss function \eqref{eq: hybrid a}, as illustrated by the yellow arrows of Fig. \ref{fig: se2e framework}.

Reversely, in the coordinated differential \textit{backward} propagation, the third layer calculates and back-propagates the gradients of the proposed hybrid loss function with respect to the var decisions, introduced in section \ref{sub: evaluate} and \ref{sub: hybrid}; the second layer calculates and back-propagates the gradients of the var decision with respect to the multiple predictions based on the third layer gradients, introduced in section \ref{sub: socp map}; the first layer calculates/back-propagates the gradients of the predictions with respect to the NN parameters $\theta_i$ and then updates $\theta_i$ of each prediction model based on the second layer gradients, as illustrated by the blue arrows of Fig. \ref{fig: se2e framework}. 

In the {\rv above} proposed framework, the decision quality from multiple predictions is measured for training prediction models reversely. The learning processes of multiple prediction models are coupled through the SOCP-driven voltage regulation model.

As the forward pass is explicit by stacking \eqref{eq: PV prediction model}, \eqref{eq: voltage regulation opt}, and \eqref{eq: decision measure} sequentially, we focus on the backward pass by differentiating through the above three layers in the following parts, as illustrated in \eqref{eq: backward total}.

\begin{subequations}
  \allowdisplaybreaks
  \begin{align}
    & \mathsf{D}_{\theta} {\mathcal{L}^{hybrid}(\hat{q}, p)} = \mathsf{D}_{\hat{q}}{\mathcal{L}^{hybrid}(\hat{q}, p)} \mathsf{D}_{\hat{p}} {\hat{q}(\hat{p})} \mathsf{D}_{\theta} {\hat{p}({\rv x})} \label{eq: backward a}\\
    &\text{Differentiate via decision-evaluating: } \mathsf{D}_{\hat{q}}{\mathcal{L}^{hybrid}(\hat{q}, p)} \label{eq: backward b}\\
    &\text{Differentiate via decision-making: } \mathsf{D}_{\hat{p}} {\hat{q}(\hat{p})} \label{eq: backward c}\\
    &\text{Differentiate via multiple prediction: } \mathsf{D}_{\theta} {\hat{p}({\rv x})} \label{eq: backward d}
  \end{align}
  \label{eq: backward total}
\end{subequations}
where $\hat{p}$ and $\theta$ are the predictions and parameters of PV/load prediction models for simplicity; $\hat{q}$ are the regulation decisions under the predicted values $\hat{p}$. We note that \eqref{eq: backward b} and \eqref{eq: backward c} are in a SOCP-driven coordinated differential way, while \eqref{eq: backward d} is in an NN-driven individual differential way.

\subsection{Decision Evaluating}\label{sub: evaluate}

The decision-evaluating layer receives the voltage regulation decisions $\hat{q}^{reg}$ from the parameterized SOCP-driven decision model under $\hat{p}:=(\hat{p}^{PV}_i, \hat{p}^{D}, \hat{q}^{D})$, calculates the total power loss under the predicted decisions $\hat{q}^{reg}$ and the ground truth PV/load outputs ${p}:=({p}^{PV}_i, {p}^{D}, {q}^{D})$, proposes the \textit{regulation loss} with the weighted sum of the power loss for economy and the voltage violation penalty for safety awareness in \eqref{eq: reg loss}, and takes minimizing the \textit{regulation loss} as the objective of the proposed decision model, which is formulated in a parameterized SOCP problem in \eqref{eq: decision measure}. 

\begin{equation}
  \mathcal{L}^{reg} (\hat{q}, p) = \underbrace{\mathcal{L}^{loss}(\hat{q}, p)}_{\text{Power loss}}  + \lambda \underbrace{\mathcal{L}^{penalty} (\hat{q}, p)}_{\text{Voltage violation}}.
  \label{eq: reg loss}
\end{equation}

\eqref{eq: voltage regulation opt} does not define the loss function explicitly, so we transform its objective to an equivalent epigraph form in the following \eqref{eq: decision measure}. 

\begin{subequations}
  \small
  \allowdisplaybreaks
  \begin{align}
    \min_{\mathcal{L}^{loss}} & \mathcal{L}^{loss} + \lambda \mathcal{L}^{penalty} \label{eq: decision measure a}\\
    \text{s.t. } & \mathcal{L}^{loss} \geq \sum_{t\in T} \sum_{ij \in E} l_{ij,t} r_{ij} \label{eq: decision measure b}\\
    & \mathcal{L}^{penalty} = \text{ReLU}(v-(V^{max}_{i,t})^2) + \text{ReLU}((V^{min}_{i,t})^2-v) \\
    & \sum_{jk \in E} P_{jk,t} - \sum_{ij \in E} (P_{ij, t} - r_{ij}l_{ij,t}) = p^{PV}_{j,t} - p^{D}_{j,t} \\
    & \sum_{jk \in E} Q_{jk,t} - \sum_{ij \in E} (Q_{ij, t} - x_{ij}l_{ij,t}) = \hat{q}^{reg}_{j,t} - q^{D}_{j,t} \\
    & v_{j,t} = v_{i,t} + (r^2_{ij}+x^2_{ij})l_{ij,t} - 2 (r_{ij}P_{ij,t} + x_{ij}Q_{ij,t}) \\
    & || 2P_{ij,t} \quad 2Q_{ij,t} \quad l_{ij,t}-v_{i,t} || \leq l_{ij,t} + v_{i,t}
  \end{align}
  \label{eq: decision measure}
\end{subequations}
where $\text{ReLU}(x)$ is the rectified linear unit, defined as $ max\{0, x \}$.

\eqref{eq: decision measure a} and \eqref{eq: decision measure b} formulate the epigraph of the original convex function, which is a convex region. \eqref{eq: decision measure} takes the $\hat{q}^{reg}$ and $p$ as input parameters and takes minimizing the generated $\mathcal{L}^{reg}$ as the learning objective. We note that lower $\mathcal{L}^{reg}$ implies higher decision quality. If the multiple predictions are accurate, the power loss $\mathcal{L}^{reg}$ is the lowest due to the optimality of the convex problem. However, the intrinsic stochasticity of PV and load determine the existence of prediction errors. We propose the power loss regret to measure the economy of decision quality numerically, as follows:

\begin{definition}[Regret of Power Loss]\label{def: regret}
  Regret of power loss defines the difference between the power loss under the predicted decisions from the semi-end-to-end decision model and that under the oracle decisions in \eqref{eq: regret}.
\end{definition}

\begin{equation}
  \text{Regret} := \mathcal{L}(\hat{q}^{reg}, p) - \mathcal{L}(q^{reg}, p)
  \label{eq: regret}
\end{equation}
where $\mathcal{L}(q^{reg}; p)$ is the optimal power loss under the oracle decisions $q^{reg}$.

\subsection{The Solution Map with Its Derivative of SOCP}\label{sub: socp map}

\begin{figure}[ht]
  \centering
  \includegraphics[scale=0.85]{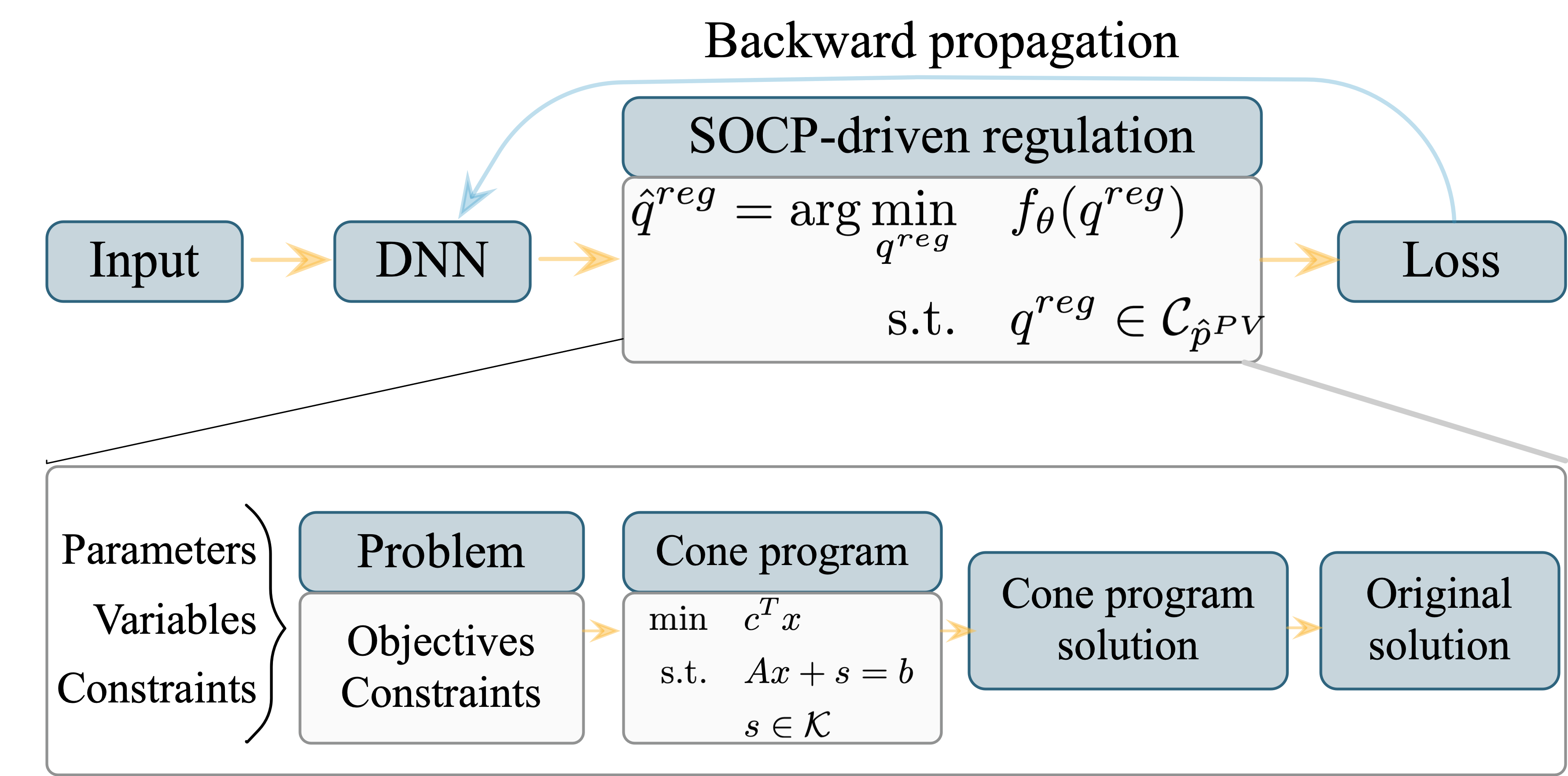}
  \caption{The solution map and its derivative of the general SOCP.}
  \label{fig: cvxlayers}
\end{figure}

The keys of \eqref{eq: backward total} back-propagation lie in $\mathsf{D}_{\hat{q}}{\mathcal{L}^{hybrid}(\hat{q}, p)}$ (\ref{eq: backward b}) and $\mathsf{D}_{\hat{p}} {\hat{q}(\hat{p})}$ (\ref{eq: backward c}). For \eqref{eq: backward c}, the parameterized SOCP-driven regulation model maps the input parameters $\hat{p}$ to the optimal decision $\hat{q}$ by the SOCP (a type of convex cone program), which does not admit the analytical solution between the $\hat{p}$ and $\hat{q}$. But it is possible to differentiate through the SOCP problem by implicitly differentiating its optimality conditions. The main scheme of the SOCP solution map and its derivative is presented in Fig. \ref{fig: cvxlayers}.

The parameterized SOCP is decomposed into three stages: an affine map from input parameters to skew matrix, a solver, and an affine map from the solver solution to the original problem solution. Then the gradients from the solution to the input parameters can be back-propagated based on the chain rule. The rest elaborates on the above three components under a general convex conic program formulation.

The general form of the SOCP-driven voltage regulation model (\ref{eq: voltage regulation opt}) is formulated as a conic program with the primal and dual forms as follows:
\begin{equation}
  \begin{aligned}[c]
    (\text{Primal}) \min \text{ }  & c^T x \\
    \text{s.t.} \text{ } & A x + s = b \\
    & s \in \mathcal{K} \\
  \end{aligned}
  \quad
  \begin{aligned}[c]
    (\text{Dual}) \min \text{ } & b^T y \\
    \text{s.t.} \text{ } & A^T y + c = 0 \\
    & y \in \mathcal{K}^* 
  \end{aligned}
  \label{eq: general socp}
\end{equation}
where $x\in\mathbb{R}^n$ and $y\in\mathbb{R}^m$ are the primal and dual variables; $s$ is the primal slack variable; $\mathcal{K}\in\mathbb{R}^m$ is the closed, convex cone with its dual cone $\mathcal{K}^*\in\mathbb{R}^m$, including the second-order cone constraints and imbalance equations of \eqref{eq: decision measure}.

Then the optimal $(x^*, y^*, s^*)$ of \eqref{eq: general socp} satisfies the Karush-Kuhn-Tucker (KKT) conditions as follows:
\begin{subequations}
  \begin{align}
    Ax + s = b \\
    A^T y + c = 0 \\
    s \in \mathcal{K}, y \in \mathcal{K}^* \\
    s^T y = 0 .
  \end{align}
  \label{eq: optimal condition}
\end{subequations}

We define the solution mapping from $(A, b, c)$ to the optimal solution $(x^*, y^*, s^*)$ as $\mathcal{S}: \mathbb{R}^{m\times n} \times \mathbb{R}^m \times \mathbb{R}^n \to \mathbb{R}^{n+2m}$ and then decompose the $\mathcal{S}$ into $\phi \circ s \circ Q$, where $\circ$ is the function composition operator that $h = f \circ g$ refers to $h(x) = f(g(x))$. Specifically, $Q$ maps the input parameters to the corresponding skew-symmetric matrix $Q$, detailed in the following part (a); $s$ solves the homogeneous self-dual embedding for implicit differentiation with the intermediate variable $z$, detailed in part (b); $\phi$ maps the $z$ to the optimal primal-dual solution in part (c).

Based on the composition, the corresponding total derivative of the solution map $\mathcal{S}$ is derived as: 
\begin{equation}
  \mathsf{D} \mathcal{S} (A, b, c) = \mathsf{D} \phi (z) \mathsf{D} s(Q) \mathsf{D} Q(A, b, c).
  \label{eq: solution map}
\end{equation}

\paragraph{Skew-symmetric Mapping}\label{para: D a}
The skew-symmetric mapping is formulated to reorganize the input parameters as:
\begin{equation}
  Q = Q(A, b ,c) =
  \begin{bmatrix}
    0 & A^T & c \\
    -A & 0 & b \\
    -c^T & -b^T & 0 \\
  \end{bmatrix} \in \mathcal{Q}.
  \label{eq: Q matrix}
\end{equation}

\paragraph{Homogeneous Self-dual Embedding}\label{para: D b}

The homogeneous self-dual embedding in Ref.\cite{Ye1994} solves \eqref{eq: general socp} by finding a zero point of a certain residual map. The embedding utilizes the variable $z$ as an intermediate variable, partitioned as $(u, v, w)$. So the normalized residual map in \cite{Busseti2019} is defined as:
\begin{equation}
  \mathcal{N}(z, Q) = ((Q - I)\Pi + I) (z / |\omega|)
  \label{eq: homogeneous}
\end{equation}
where $\Pi$ is the project operation onto $\mathbb{R} \times \mathcal{K} \times R_{+}$.

If and only if $\mathcal{N}(z, Q) = 0$ and $\omega >0$, the $z$ can construct the solution of \eqref{eq: general socp} for given $Q$. Then we derive the derivatives of $\mathcal{N}(z, Q)$ with respect to $z$ and $Q$, respectively in \eqref{eq: DQN} and \eqref{eq: DzN}.

\begin{equation}
  \mathsf{D}_{Q} \mathcal{N}(z, Q) = \Pi(z/|\omega|).
  \label{eq: DQN}
\end{equation}
\begin{subequations}
  \begin{align}
    \mathsf{D}_{z} \mathcal{N}(z, Q) 
    =& ((Q-I)\mathsf{D}\Pi(z)+I)/\omega \nonumber \\
    &- \text{sign}(\omega)((Q-I)\Pi + I)(z/\omega^2) e^T  \\
    =& ((Q-I)\mathsf{D}\Pi(z)+I)/\omega \\
    &\text{ when $z$ is the solution of \eqref{eq: general socp}} \nonumber
  \end{align}
  \label{eq: DzN}
\end{subequations}
where $e \in \mathbb{R}^n$ is $(0,0, ..., 1)$. 

According to the implicit function theorem \cite{Dontchev2009}, there exists a neighborhood $V \subseteq \mathcal{Q}$ such that the $z=s(Q)$ of $\mathcal{N}(z, Q)$ is unique. So $\mathsf{D}_Q\mathcal{N}(z(Q), Q)$ is zero when $z$ is the solution of \eqref{eq: general socp}. Then the derivative of $s(Q)$ with respect to Q is derived as:

\begin{subequations}
  \begin{align}
    & \mathsf{D} \mathcal{N}_{z}(s(Q, Q) \mathsf{D} s(Q) +  \mathsf{D}_{Q} \mathcal{N}(z, Q) = 0 \\
    \Rightarrow & \mathsf{D} s(Q) = - \mathsf{D} \mathcal{N}_{z}(s(Q, Q)^{-1} \mathsf{D}_{Q} \mathcal{N}(z, Q).
  \end{align}
  \label{eq: DsQ}
\end{subequations}

\paragraph{Solution Construction}\label{para: D c}

The solution construction function of mapping $z=(u, v, \omega) \in \mathbb{R}^N$ to $(x^*, y^*, s^*)\in\mathbb{R}^{n+2m}$ is defined as:
\begin{equation}
  \phi(z) = (u, \Pi_{\mathcal{K}^*}(v), \Pi_{\mathcal{K}^*}(v)) / \omega.
  \label{eq: solution constr}
\end{equation}
So the $\phi(z)$ is differentiable with derivative as

\begin{equation}
  \mathsf{D} \phi (z) =
  \begin{bmatrix}
    I & 0 & -x \\
    0 & \mathsf{D} \Phi_{\mathcal{K}^*}(v) & -y \\
    0 & \mathsf{D} \Phi_{\mathcal{K}^*}(v) - I & -s \\
  \end{bmatrix}.
\end{equation}

So the derivative of $\mathsf{D}\mathcal{S}(A, b, c)$ is formulated by stacking the above three derivative components:
\begin{equation*}
  \mathsf{D} \mathcal{S} (A, b , c) = 
  \underbrace{\mathsf{D} \phi (z)}_{\substack{\text{Solution} \\ \text{construction} \\ \text{derivative}}} 
  \underbrace{\mathsf{D} s(Q)}_{\substack{\text{Homogeneous} \\ \text{self-dual embedding}\\ \text{derivative}}} 
  \underbrace{\mathsf{D} Q(A, b, c)}_{\substack{\text{Skew-symmetric} \\ \text{mapping} \\ \text{derivative}}}.
\end{equation*}

The above implicit derivative of the SOCP solution calculation method (\ref{eq: solution map}) can obtain the backward propagation derivatives of \eqref{eq: backward b} and \eqref{eq: backward c} in the coordinated differential way. \eqref{eq: backward d} can be calculated by the automatic differentiating algorithm of PyTorch \cite{Paszke2019}. So the total derivative of the decision objective with respect to the multiple prediction model parameters can be obtained through the chain rule of composite function in \eqref{eq: backward total}. In this way, the impact on voltage regulation from multiple PV and load predictions is integrated into the total derivatives, coordinating the prediction models for enhancing decision quality.

\subsection{Hybrid Training Strategy Considering Prediction Accuracy and Decision Quality}\label{sub: hybrid}

Based on the evaluating of prediction accuracy (\ref{eq: mse training}) and decision quality (\ref{eq: decision measure}), this paper proposes the hybrid loss in \eqref{eq: hybrid a} based on the coordinated \textit{regulation loss} and the individual prediction loss; the gradients of prediction errors \eqref{eq: solution map} and decision objective \eqref{eq: mse derivative} in \eqref{eq: hybrid b} are derived and combined for hybrid training the multiple prediction models.

\begin{subequations}
  \begin{align}
    &\mathcal{L}^{hybrid}_{i} = \mathcal{L}^{reg}+ \epsilon \mathcal{L}^{mse}_i \quad \forall i \in \mathcal{N}_{pred}\label{eq: hybrid a}\\
    \Rightarrow &\mathsf{D}_{\theta_i} \mathcal{L}^{hybrid} = \underbrace{\mathsf{D}_{\theta_i} \mathcal{L}^{obj}}_{\text{By \eqref{eq: solution map}}} + \epsilon \underbrace{\mathsf{D}_{\theta_i} \mathcal{L}^{mse}_i}_{\text{By \eqref{eq: mse derivative}}}  \quad \forall i \in \mathcal{N}_{pred}\label{eq: hybrid b}
  \end{align}
  \label{eq: hybrid gradient}
\end{subequations}
where \eqref{eq: hybrid a} defines $\mathcal{L}^{hybrid}_{i}$ as the hybrid loss function for prediction model $i$; \eqref{eq: hybrid b} calculates the gradients of the hybrid loss with respect to prediction model parameters; $\mathcal{N}_{pred}$ is the set of multiple PV/load prediction models.

So this paper proposes the safety-aware semi-end-to-end hybrid stochastic gradient descent (HSGD) learning algorithm for training multiple PV and load prediction models, as illustrated in algorithm \ref{algo: sgd algo}. {\rv The proposed HSGD is based on the Adam optimizer for utilizing the gradients from the regulation model to update the parameters of neural networks. Adam is a first-order gradient-based optimization algorithm of stochastic objective functions with efficient convergence performance \cite{Kingma2015}.} Firstly, it initializes with the multiple PV datasets, essential information of voltage regulation model (\ref{eq: voltage regulation opt}), hyperparameters, and pre-trained the prediction models based on the MSE in \eqref{eq: mse loss} to avoid the infeasibility in the decision-making layer. Then it maps the input multiple input features to the hybrid loss through the predicting, decision-making, and decision-evaluating in the coordinated forward pass, calculates the derivatives of MSE and \textit{regulation} loss function with respect to the $\theta_i$ as $\mathsf{D}_{\theta_i} \mathcal{L}^{reg}$ and $\mathsf{D}_{\theta_i} \mathcal{L}^{mse}_{i}$ in the coordinated differential backward pass, and updates each prediction model parameters based on the weighted sum gradients by gradient descent. Finally, the trained semi-end-to-end decision model can be applied for online decisions. {\rv It should be noted that though the integration of the SOCP-driven model will impose a higher calculation burden on offline training ML models than the convention MSE-based model, it will not influence the practical online implementation of the proposed model. Because the day-ahead scheduling does not have the real-time calculation requirement, and online prediction of trained models is quite fast.}

\begin{algorithm}[ht]
  {\small{
    \begin{algorithmic}[1]
    \STATE \textbf{Input:} Total Dataset $\mathcal{D}=\{(x_i, p^{pred}_i)\}_{i \in \mathcal{N}_{pred}}$ {\rv where the input context $x_i$ is composed of the past day power and related NWP information}, the voltage regulation model information, pre-trained epoch $N_{pre}$, training epoch $N_{train}$, learning rate $\alpha$, and weight $\epsilon$;
    \STATE \textbf{Initialize:} Prediction model set $\mathcal{F} = \{f^{pred}_{\theta_i}(\cdot)\}_{i \in \mathcal{N}_{pred}}$, pre-training prediction models based on prediction error by \eqref{eq: mse training}, and separate the $\mathcal{D}$ into $\mathcal{D}_{train}$ and $\mathcal{D}_{test}$;
    \FOR{Epoch = 1, ..., $N_{train}$}
    \STATE \textbf{Coordinated forward pass}
    \STATE \textit{Predicting}: Make multiple PV and load predictions from different context $x_i$ and obtain $\{\hat{p}^{pred}_i = f^{pred}_{\theta_i}(x_i)\}_{i\in \mathcal{N}_{pred}}$;
    \STATE \textit{Decision-making}: Stack multiple predictions and deliver them to parameterized SOCP-driven voltage regulation model (\ref{eq: voltage regulation opt}) to make decisions $\hat{q}^{reg}$;
    \STATE \textit{Decision-evaluating}: Evaluate the decision quality of var regulation decisions $\hat{q}^{reg}$ under the true PV power $p^{pred}$ by $\mathcal{L}^{reg}(\hat{q}^{reg}, p^{pred})$ from \eqref{eq: decision measure};
    \STATE \textbf{Coordinated differential backward pass}
    \STATE Calculate the derivative of \textit{mse} with respect to $\theta_i$ as $\mathsf{D}_{\theta_i} \mathcal{L}^{mse}_i$ for $i \in \mathcal{N}_{pred}$;
    \STATE Calculate the derivative of decision objective with respect to $\theta$ as $\mathsf{D}_{\theta} \mathcal{L}^{reg}$, where $\theta = (\theta^T_1, ... ,\theta^T_{N_{pred}})^T$;
    \STATE \textbf{Gradient step}
    \STATE Run gradient descent to update prediction models: 
    $$\theta_i \leftarrow \theta_i - \alpha (\mathsf{D}_{\theta_i} \mathcal{L}^{reg}+ \epsilon \mathsf{D}_{\theta_i} \mathcal{L}^{mse}_i)$$
    \ENDFOR
    \STATE {\rv\textbf{Output:} Trained semi-end-to-end coordinated decision model.}
    \hfill
    \STATE \textbf{Online Decision-making:}
    \STATE {\rv Input} PV/load contexts $x_i$ {\rv with the past day power and related NWP information}, construct $\{\hat{p}^{PV}_i = f^{PV}_{\theta_i}(x_i)\}_{i\in \mathcal{N}_{PV}}$ based on NN, and {\rv output voltage regulation decisions $\hat{q}^{reg}$} based on SOCP model.
    \end{algorithmic}
  }
  }
  \caption{Safety-aware Semi-end-to-end HSGD Learning Algorithm for Voltage Regulation.}
  \label{algo: sgd algo}
\end{algorithm}

\section{Case Study}\label{sec: case}

\begin{figure}[ht]
  \centering
  \includegraphics[scale=1]{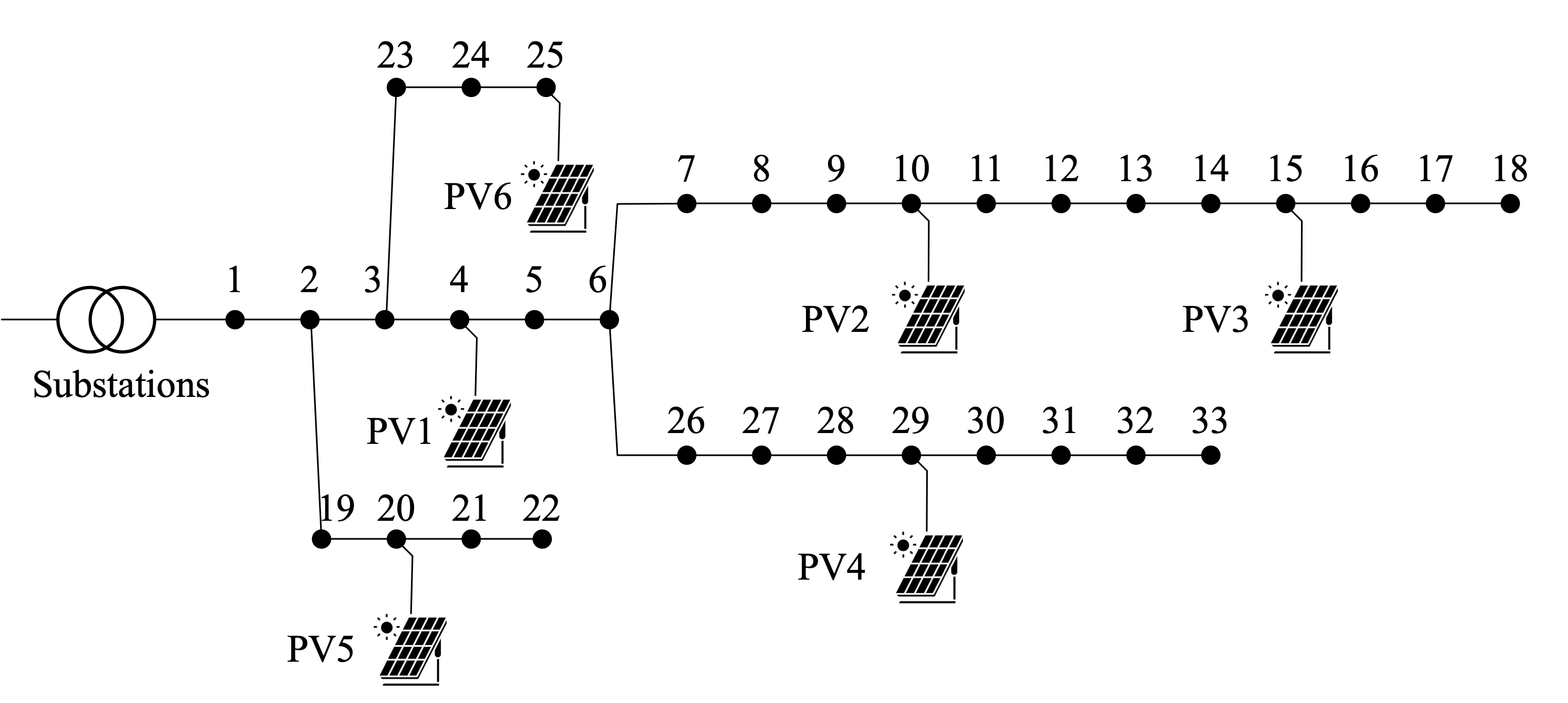}
  \caption{Network topology and PV locations.}
  \label{fig: case 33}
\end{figure}

We utilize a 33-bus radial distribution network \cite{Baran1989} to verify the economy and safety awareness of the proposed semi-end-to-end coordinated decision model. Multiple PV station data combined with related weather information from 2018 to 2019 are utilized for training the semi-end-to-end model for voltage regulation from {\rv the} State Grid {\rv Corporation of China} Competition, and the corresponding load data are from PJM market \cite{PJM}. The network topology, inverter-based regulator location, and PV systems location are presented in Fig. \ref{fig: case 33}. The proposed decision model is implemented by Python 3.6 with PyTorch package \cite{Paszke2019} for the NN-driven prediction models and the Cvxlayer package \cite{Agrawal2019-1} for the voltage regulation models of \eqref{eq: voltage regulation opt}, \eqref{eq: decision measure}, which are deployed on a DELL server with CPU Intel Xeon, GPU Nvidia GTX3090, and RAM 64 GB. The necessary data for this paper are attached in Ref. \cite{Dataset}.

This section {\rv firstly} preprocesses the multiple PV/load datasets, {\rv then} initializes the prediction models by pre-training, and {\rv finally} evaluates the performance of the proposed safety-aware semi-end-to-end model from the perspectives of power loss and safety awareness. 

\subsection{Datasets and Models}

\subsubsection{Data Preprocessing}

This paper divides the raw dataset into the training set (80\%) for model training and the testing set (20\%) for model evaluation. The testing set is further split into six cases for independent study. Then it processes the data as follows: 1) normalize the multiple PV/load raw data into the scale of 0-1 for prediction model training, 2) multiply the predicted normalized power with the corresponding PV/load capacity, 3) and deliver the recovered power to downstream SOCP-driven voltage regulation for model training and decision-making. The input features of PV prediction models are the PV {\rv generation} of the past day and the predicted temperature/irradiance of the predicted day from {\rv the NWP} information; the input features of the load prediction model are the load demand of the past day and the predicted temperature of the predicted day. This paper utilizes five SVCs with 500 kVar in buses 6, 13, 14, 29, and 31 for voltage regulation. {\rv It should be noted that the prediction models consider the seasonal pattern of PV generation and load demand by taking the NWP data as input, which are seasonal-adaptive and can be applied to non-stationary scenarios for the practical distribution network.}

\begin{table}[ht]
  \renewcommand{\arraystretch}{1.3}
  \centering
  \caption{PV location and capacities.}
  \begin{tabular}{ccc|ccc}
    \hline
    PV No. & Bus & Capacity/kW & PV No. & Bus & Capacity/kW \\ \hline
    1 & 4 & 600 & 4 & 29 & 300 \\
    2 & 10 & 300 & 5 & 20 & 300 \\
    3 & 15 & 400 & 6 & 25 & 700 \\ \hline
    \end{tabular}
  \label{tab: capacity}
\end{table}

\subsubsection{Model Initializing} 

\begin{figure}[ht]
  \centering
  \rv 
  \subfigure[Multiple PV initializing.]{\includegraphics[scale=0.95]{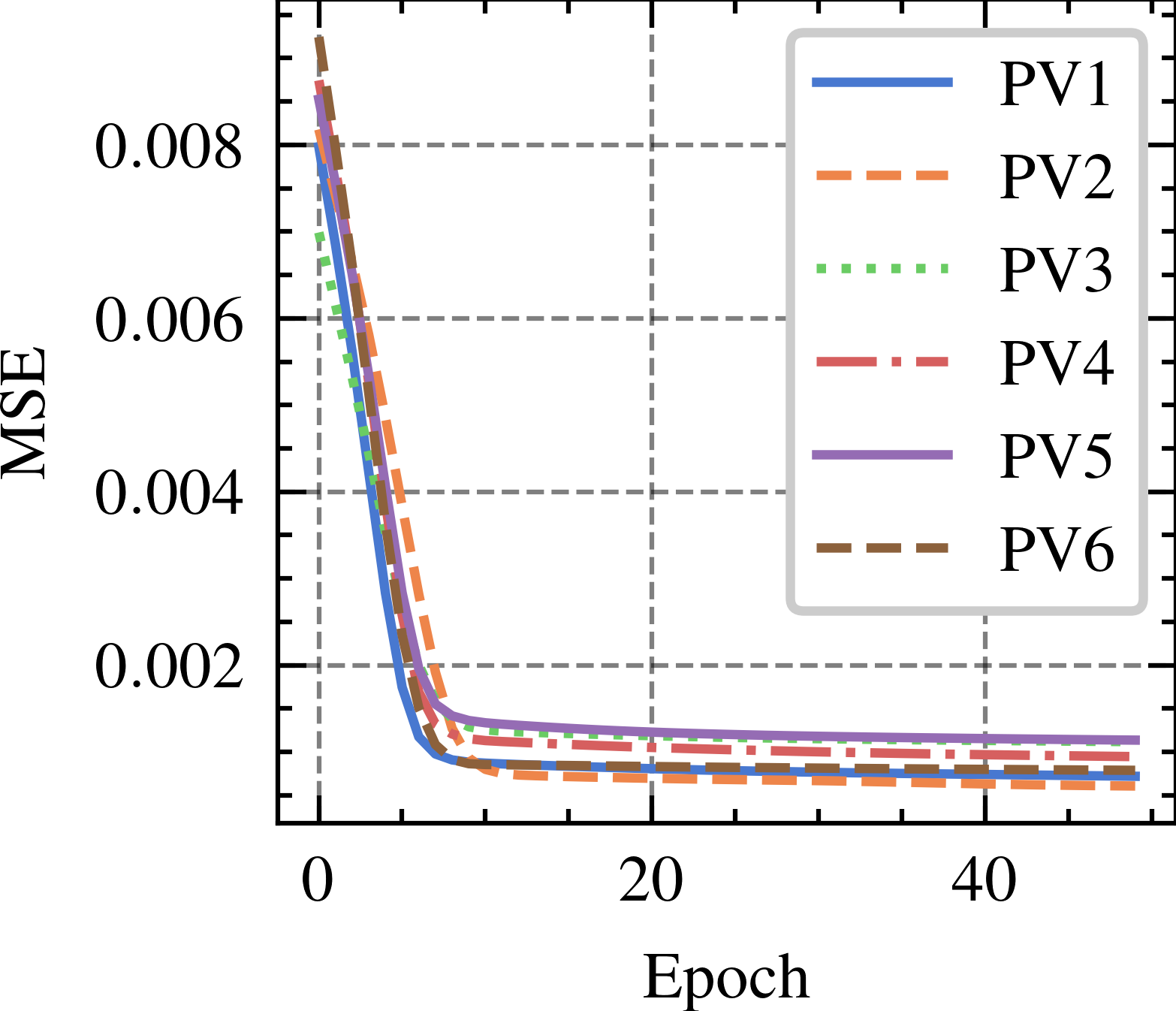}}
  \subfigure[Multiple load initializing.]{\includegraphics[scale=0.95]{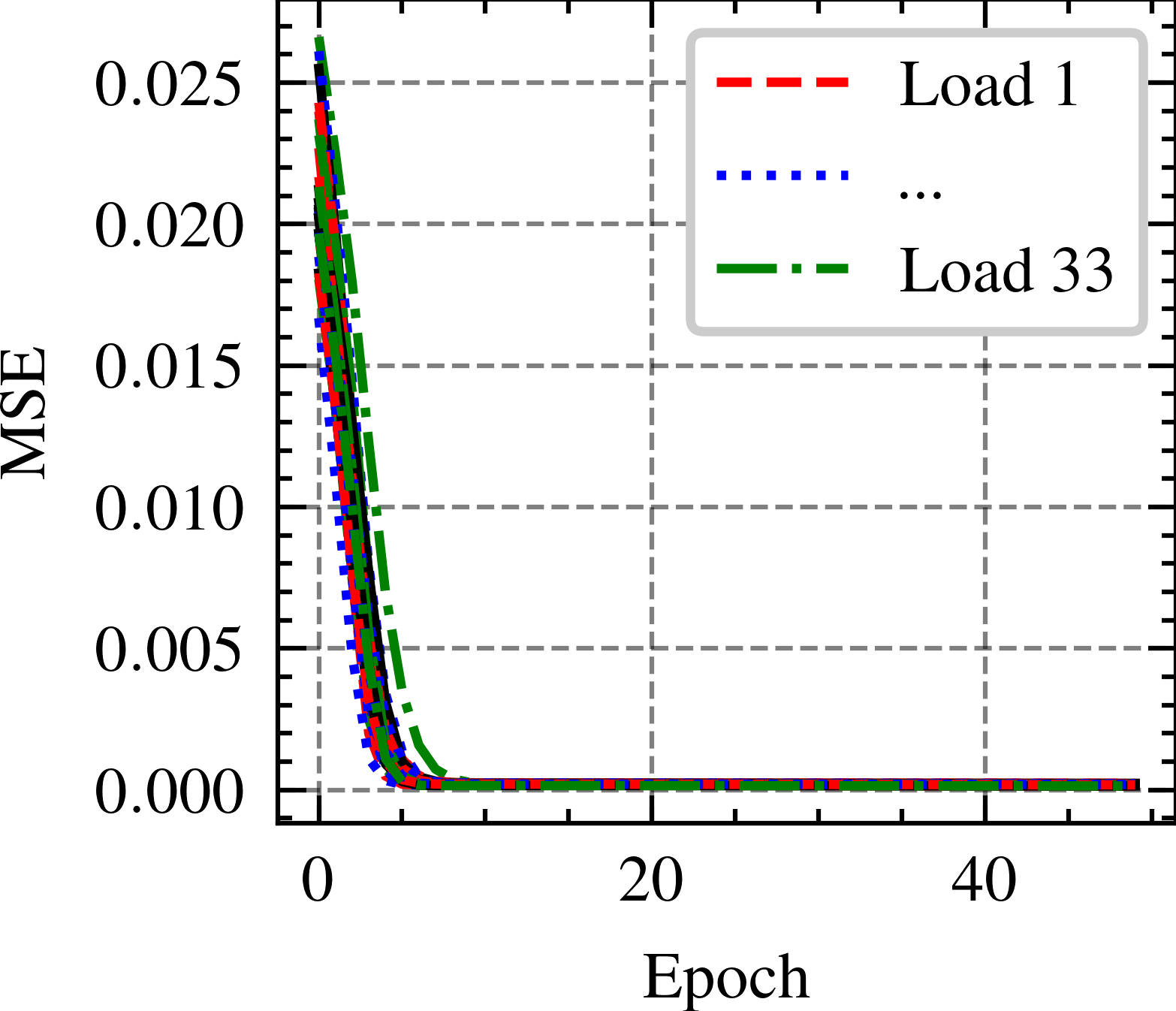}} 
  \caption{Prediction initializing process.}
  \label{fig: initial}
\end{figure}

Then we initialize the multiple prediction models by conventional MSE-based training and then further train the above models by the proposed semi-end-to-end HSGD learning algorithm \ref{algo: sgd algo}. The initial high inaccuracy of predictions may lead to the infeasibility of downstream optimization models and further result in a longer training time. Fig. \ref{fig: initial} shows the change of \textit{mse} in the model initializing process {\rv with multiple PV and load} and demonstrates that the predictions models converge after ten epochs. 

The corresponding semi-end-to-end model hyperparameter values are presented in Tab. \ref{tab: nn param}.

\begin{table}[ht]
  \renewcommand{\arraystretch}{1.3}
  \centering
  \caption{Model hyperparameter values {\rv of the NN and SOCP models}.}
  \begin{tabular}{cc|cc}
    \hline
    Hyperparameters & Values & Hyperparameters & Values  \\
    \hline
    Hidden layers & (128, 128) & Optimizer & Adam \\
    Learning rate & 3 $\times$ 10$^{-4}$ & Decay rate & 1 $\times$ 10$^{-5}$ \\
    Activation function & ReLU & PV number & 6 \\
    $\epsilon$ & 1 & $\lambda$ & 1 \\
    \hline
  \end{tabular}
  \label{tab: nn param}
\end{table}

\subsection{Performance of the Semi-end-to-end Decision Model}

The objective of the decision-evaluating layer in \eqref{eq: decision measure} is composed of the power loss ($\mathcal{L}^{loss}$) and the penalty of voltage deviation ($\mathcal{L}^{penalty}$) for minimizing the power loss in a safety-aware way. However, the safety awareness of voltage regulation may sacrifice the power loss optimality. So this paper evaluates the power loss reduction and safety-aware performance of the proposed decision model in two separated scenarios, denoted by \textit{economic scenario} and \textit{safety scenario}: 1) \textit{economic scenario} focuses on the power loss reduction by broadening the allowed operational voltage as [0.94, 1.06] p.u. and doubling the value of line impedance, where the decision-making under MSE-based prediction will not violate the voltage constraints in \eqref{socp f} and the power loss improvement will be noticeable; 2) \textit{safety scenario} restricts the allowed voltage range as [0.95, 1.05] p.u., where some decision-making under the MSE-based prediction will violate the voltage constraints, and the proposed decision model can effectively address this violation problem. {\rv It should be noted that the safety scenario is the practical scenario, where both the power loss and violation penalty in the objective function take effect. Then the economic scenario is appended to exhibit the power loss reduction performance of the proposed model significantly. Economic and safety scenarios exhibit the economy and safety-awareness features of the proposed semi-end-to-end model.}

\subsubsection{Power Loss Reduction Performance}

\textit{Economic scenario} focuses on verifying the power loss reduction performance of the proposed decision model by enlarging the allowed voltage range and line impedance.

\paragraph{Training Progress}

\begin{figure}[ht]
  \centering
  \includegraphics[scale=0.85]{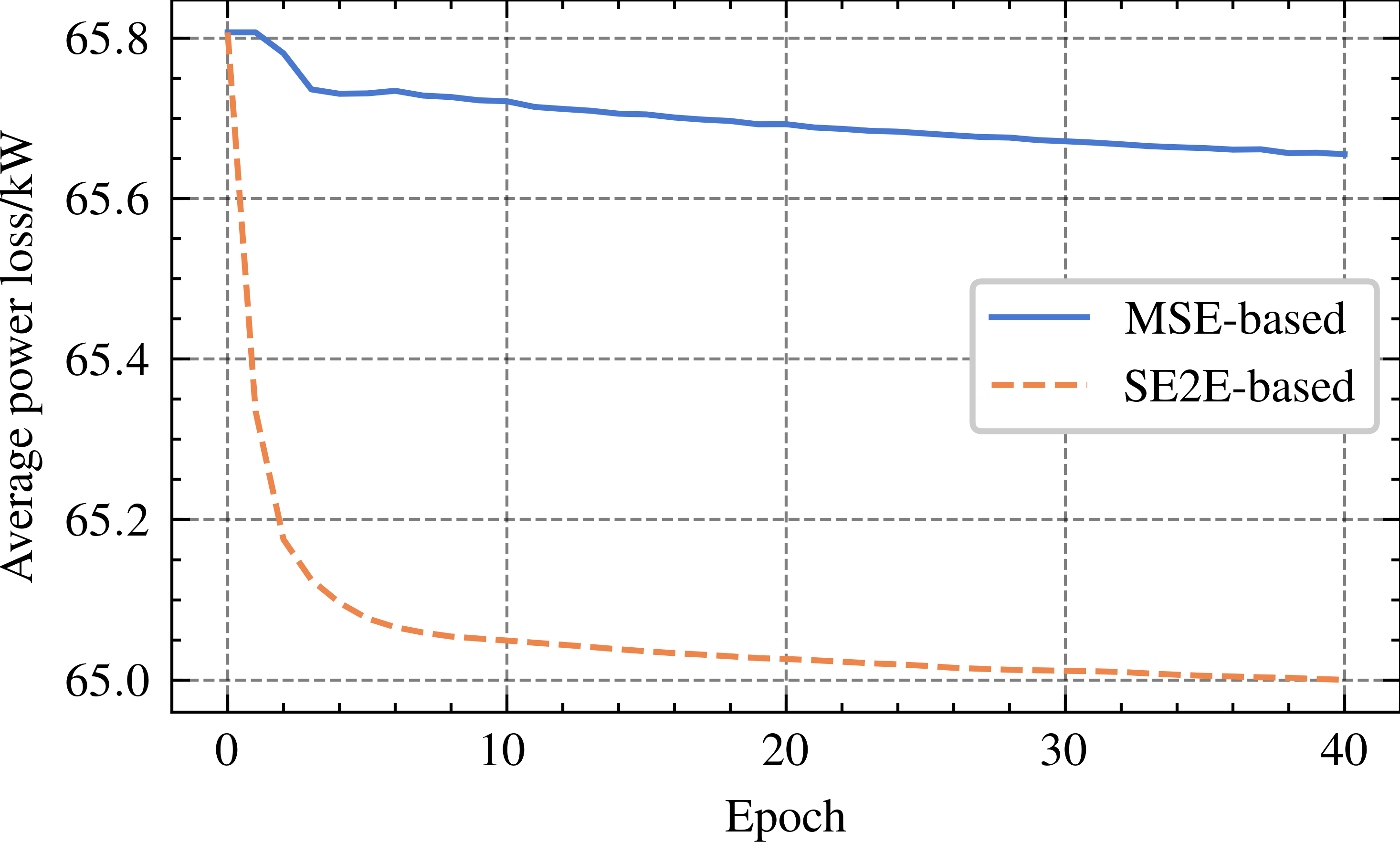}
  \caption{Power loss {\rv changing} in the training process.}
  \label{fig: pl training}
\end{figure}

Fig. \ref{fig: pl training} presents the changes of {\rv the} average power loss under the safety-aware semi-end-to-end and the MSE-based SGD learning algorithm, denoted by SE2E-based and MSE-based. The SE2E-based method reduces the power loss at the first ten epochs, verifying the effectiveness of {\rv integrating} the downstream decision-making models for further reducing the power loss.

\paragraph{Results Analysis}

\begin{figure}[ht]
  \centering
  \includegraphics[scale=0.9]{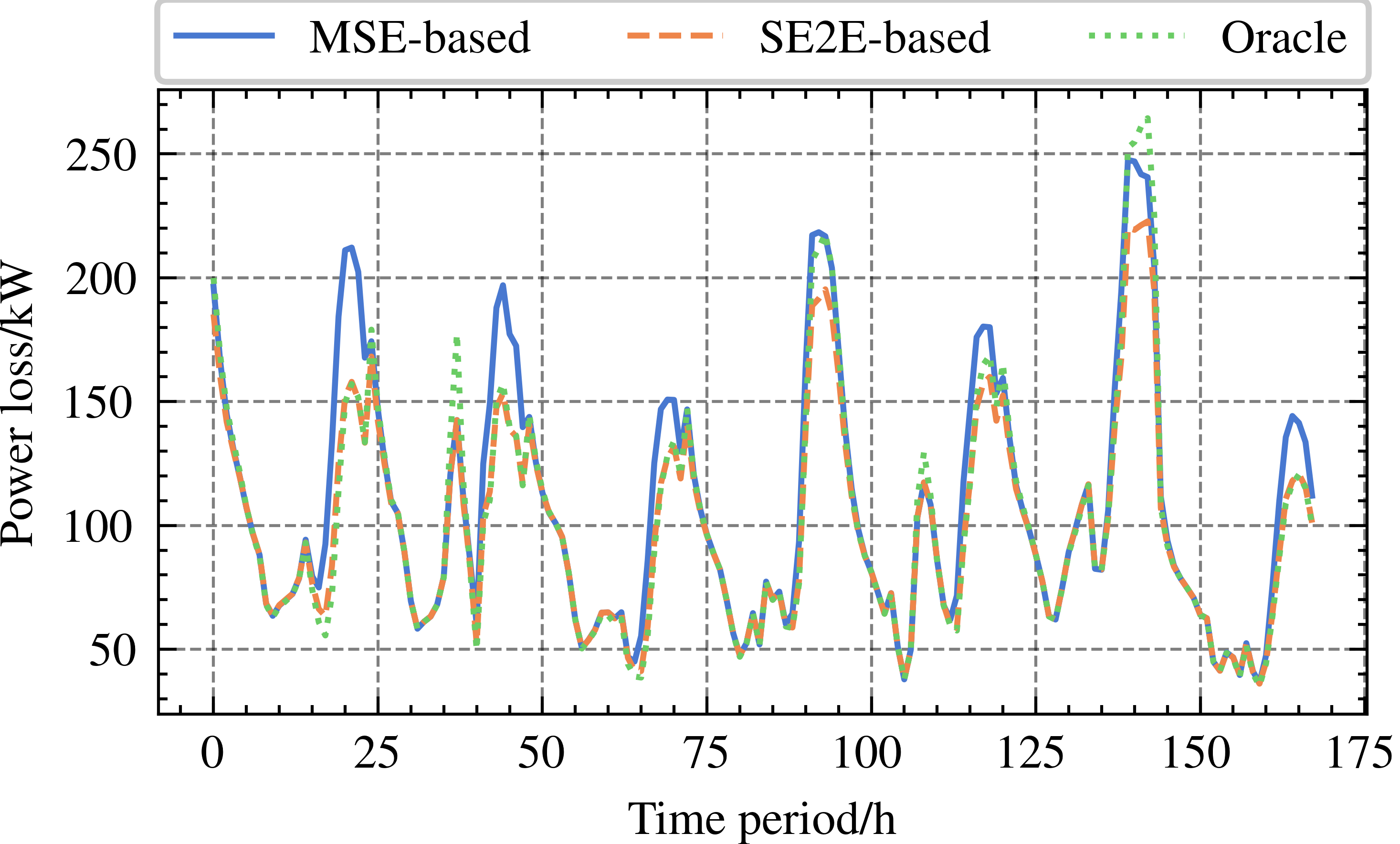}
  \caption{Power loss in a typical week.}
  \label{fig: power loss}
\end{figure}

Fig. \ref{fig: power loss} compares the power loss of a typical week under the SE2E-based, MSE-based, and oracle methods, where oracle refers to the decision-making under the true PV outputs and load demands. {\rv The power loss of the SE2E-based method is closer to that of the oracle than the conventional MSE-based method, demonstrating that the decision quality under the SE2E-based method approaches the oracle decision quality and surpasses the conventional MSE-based method.}

\begin{figure}[ht]
  \centering
  \includegraphics[scale=0.9]{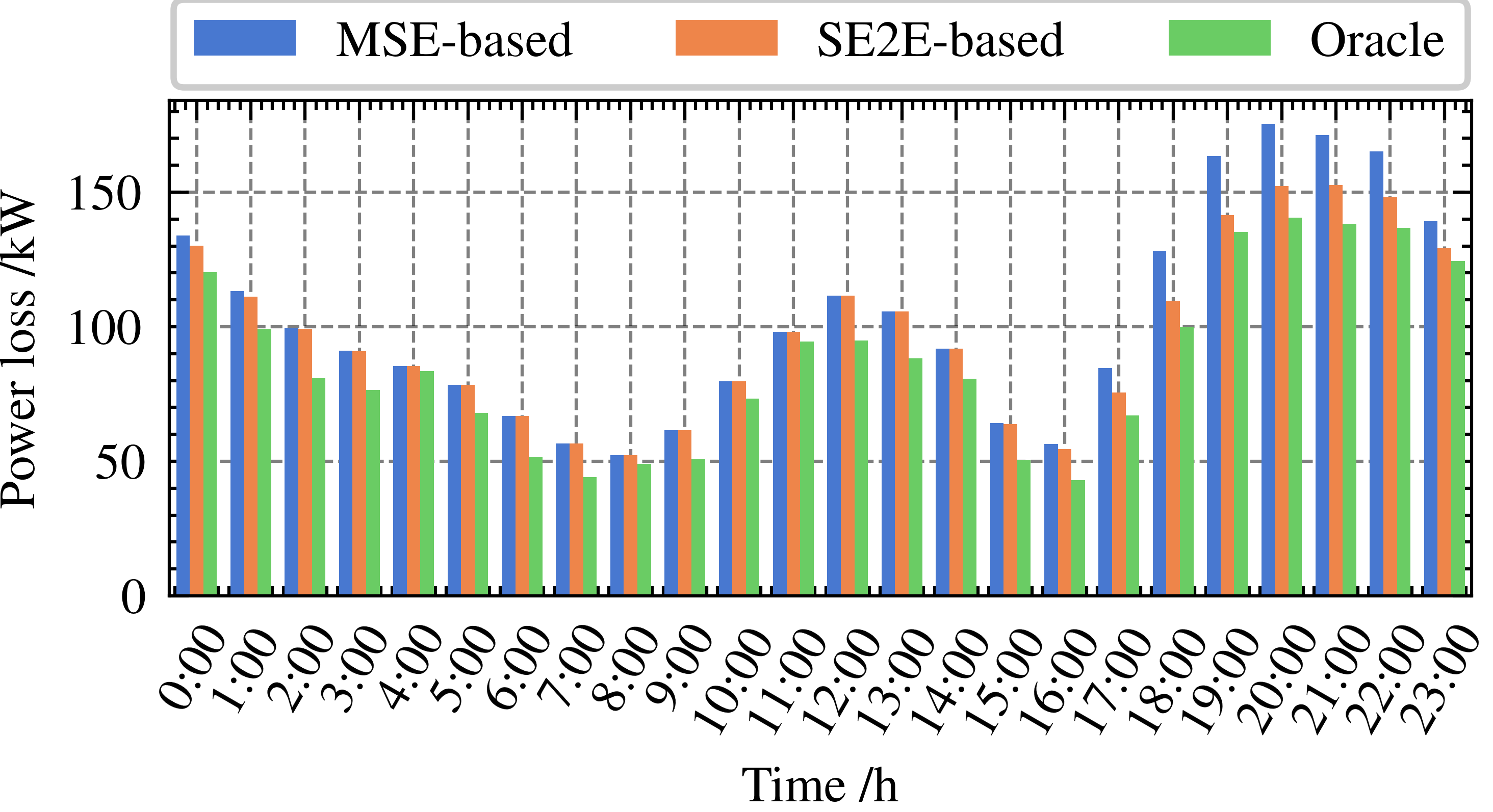}
  \caption{Power loss in different periods.}
  \label{fig: power loss avg}
\end{figure}

Fig. \ref{fig: power loss avg} further analyzes the average power loss of different periods. The MSE-based method shows high regrets than the SE2E-based method in each period. {\rv The hourly average power loss of SE2E-based is lower than that of MSE-based from 17:00 to 22:00. The above result demonstrates the power loss reduction effectiveness of the proposed decision model.}

Then we compare the prediction {\rv accuracy} and power loss reduction performance of the MSE-based and SE2E-based models numerically in the following Tab. \ref{tab: pred economic} and \ref{tab: pl economic}. The mean-absolute-mean-error (MAPE) and regret are utilized to measure the performance of prediction and decision-making.

Tab. \ref{tab: pred economic} compares the multiple prediction model MAPE under the above two models in six test cases. {\rv The multiple PV predictions under SE2E-based with lower MAPE are more accurate than those under MSE-based; in contrast, the load prediction under MSE-based shows higher accuracy.} {\rv Integrating downstream voltage regulation can improve the PV prediction accuracy but degrade some load prediction accuracy due to the objective of enhancing the decision quality.} 

\begin{table*}[ht]
  \renewcommand{\arraystretch}{1.3}
  \centering
  \caption{Prediction performance analysis in different datasets in the economic scenario.}
  \begin{tabular}{ccccccccccccccc}
    \hline
    \multicolumn{1}{c}{\multirow{2}{*}{Case}} & \multicolumn{2}{c}{MAPE of PV1} & \multicolumn{2}{c}{MAPE of  PV2} & \multicolumn{2}{l}{MAPE   of  PV3} & \multicolumn{2}{c}{MAPE of  PV4} & \multicolumn{2}{c}{MAPE of  PV5} & \multicolumn{2}{c}{MAPE of  PV6} & \multicolumn{2}{c}{MAPE of Load} \\ \cline{2-15} 
    \multicolumn{1}{c}{} & MSE & SE2E & MSE & SE2E & MSE & SE2E & MSE & SE2E & MSE & SE2E & MSE & SE2E & MSE & SE2E \\ \hline
    1 & 0.1803 & 0.1751 & 0.2810 & 0.2732 & 0.2557 & 0.2502 & 0.2794 & 0.2708 & 0.3172 & 0.3173 & 0.2239 & 0.2236 & 0.0468 & 0.0499 \\
    2 & 0.1928 & 0.1883 & 0.3012 & 0.2941 & 0.2864 & 0.2773 & 0.3067 & 0.2983 & 0.2933 & 0.2933 & 0.1944 & 0.1941 & 0.0531 & 0.0560 \\
    3 & 0.2081 & 0.2027 & 0.2683 & 0.2571 & 0.2553 & 0.2461 & 0.3164 & 0.3083 & 0.2988 & 0.2990 & 0.1986 & 0.2029 & 0.0448 & 0.0435 \\
    4 & 0.1956 & 0.1886 & 0.3008 & 0.2881 & 0.2956 & 0.2858 & 0.3078 & 0.2981 & 0.2861 & 0.2861 & 0.2104 & 0.2078 & 0.0556 & 0.0629 \\
    5 & 0.2374 & 0.2343 & 0.3379 & 0.3198 & 0.2832 & 0.2720 & 0.3137 & 0.3038 & 0.2649 & 0.2651 & 0.2060 & 0.2076 & 0.0516 & 0.0523 \\
    6 & 0.2258 & 0.2192 & 0.3078 & 0.2963 & 0.2759 & 0.2708 & 0.3096 & 0.3021 & 0.3024 & 0.3024 & 0.2351 & 0.2369 & 0.0641 & 0.0639 \\ \hline
    Avg. & 0.2067 & 0.2014 & 0.2995 & 0.2881 & 0.2754 & 0.2670 & 0.3056 & 0.2969 & 0.2938 & 0.2939 & 0.2114 & 0.2122 & 0.0527 & 0.0548 \\ \hline
    \end{tabular}
  \label{tab: pred economic}
\end{table*}

Tab. \ref{tab: pl economic} compares the power loss and regret of the above two models. The average power loss under SE2E is 0.58 kW lower than that under MSE with 0.89\% power loss reduction. SE2E decreases the decision-making regret by 52.72\% on average than MSE, verifying its loss reduction effectiveness. {\rv Though the load prediction accuracy degrades in the proposed SE2E-based model, the decision quality from multiple prediction upgrades. It demonstrates the inconsistency between prediction accuracy and decision quality.}

\begin{table}[ht]
  \renewcommand{\arraystretch}{1.3}
  \centering
  \caption{Decision performance analysis of different cases in the economic scenario.}
  \begin{tabular}{cccccc}
    \hline
    \multirow{2}{*}{Case} & \multicolumn{3}{c}{Power loss/kW} & \multicolumn{2}{c}{Regret/kW} \\ \cline{2-6}
     & MSE & SE2E & Oracle & MSE & SE2E  \\ \hline
     1 & 68.63 & 68.10 & 67.53 & 1.10 & 0.58  \\
     2 & 64.70 & 64.01 & 63.64 & 1.06 & 0.37  \\
     3 & 63.89 & 63.49 & 62.93 & 0.96 & 0.56  \\
     4 & 69.89 & 69.24 & 68.75 & 1.14 & 0.49  \\
     5 & 64.48 & 63.69 & 63.03 & 1.45 & 0.66  \\
     6 & 61.48 & 61.02 & 60.58 & 0.90 & 0.44  \\ \hline
     Avg. & 65.51 & 64.93 & 64.41 & 1.10 & 0.52 \\ \hline
    \end{tabular}
  \label{tab: pl economic}
\end{table}

{\rv
To verify the implementation effectiveness, Tab. \ref{tab: time compare} compares the mean solving time of the conventional MSE-based and the proposed SE2E-based models, and the average time of the SE2E-based is 0.17s higher than the MSE-based. This demonstrates the implementation effectiveness of the proposed model.

\begin{table}[ht]
  \renewcommand{\arraystretch}{1.3}
  \centering
  \caption{\rv Online implementation time comparison of different cases in the economic scenario.}
  \begin{tabular}{cccccccc}
    \hline
    Case & 1 & 2 & 3 & 4 &  5 & 6 & Avg. \\
    \hline
    MSE & 1.79s & 1.84s & 1.79s & 1.80s & 1.81s & 1.81s & 1.80s \\
    SE2E & 1.97s & 1.96s & 1.99 & 1.99s & 1.96s & 1.96s & 1.97s \\
    \hline
  \end{tabular}
  \label{tab: time compare}
\end{table}
}

Case study on \textit{economic scenario} demonstrates that the integration of the downstream voltage regulation model for coordinating the multiple prediction models can not only improve the decision quality but prediction accuracy of multiple PV predictions, {\rv verifying the necessity of considering the downstream decision economy.}

\subsubsection{Safety-aware Performance}

\textit{Safety scenario} focuses on verifying the {\rv safety-awareness} of the proposed decision model to address the violation problem from inaccurate decision-making. The violation rate is defined as the percentage of violation cases to the total cases, which is utilized to measure the safety performance of decision-making.

\paragraph{Training Progress}

\begin{figure}[ht]
  \centering
  \includegraphics[scale=0.85]{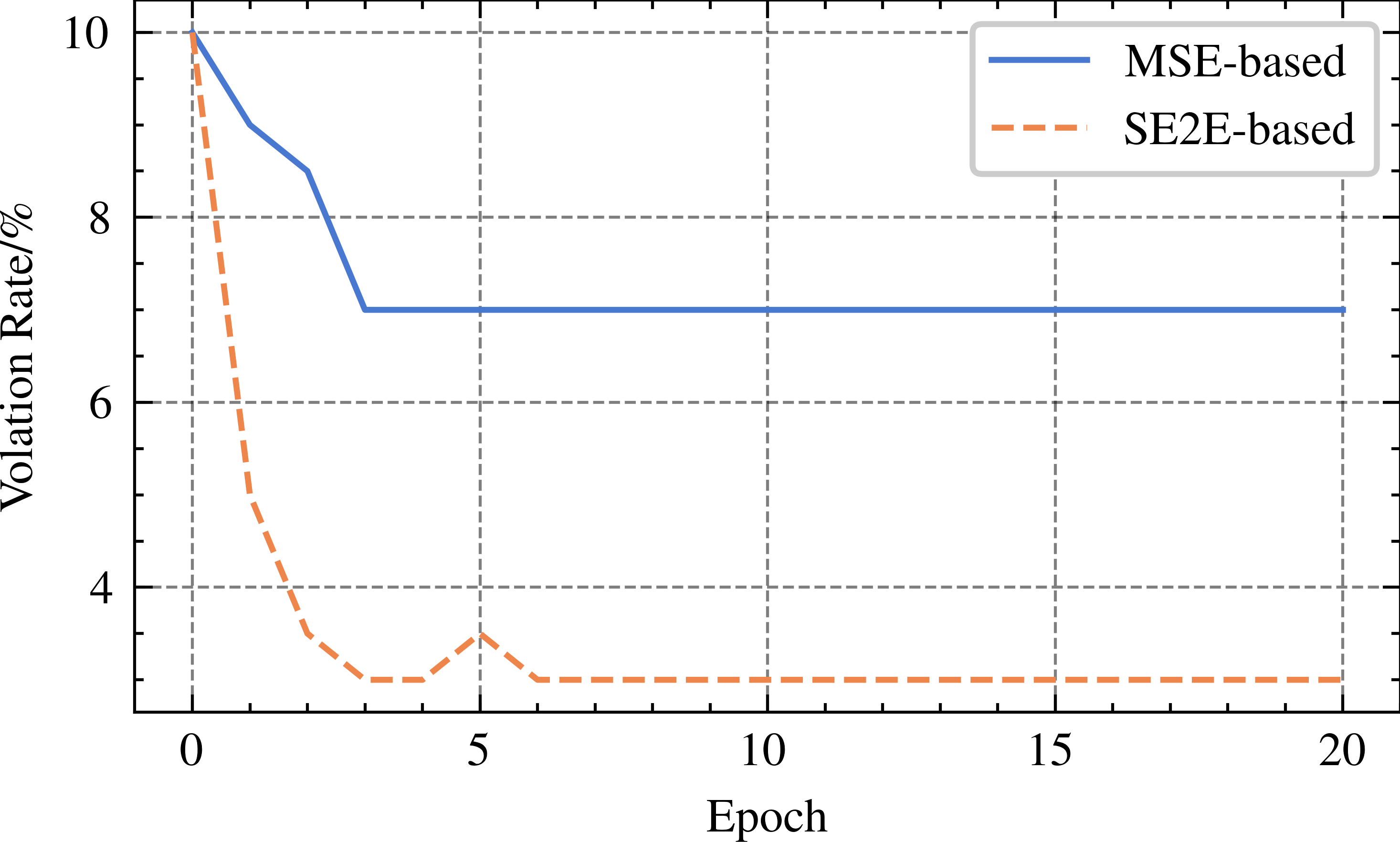}
  \caption{Violation rate changing in the process.}
  \label{fig: vio}
\end{figure}

Fig. \ref{fig: vio} presents the changes in power constraint violation rate under SE2E-based and MSE-based methods. Due to the convergence of the MSE-based method in model initializing, further training does not change the violation rate of 7\%. In contrast, the violation rate under the proposed SE2E-based method reduces to 3\% at first, then fluctuates, and finally stabilizes at 3\%, verifying the safety-aware performance of the proposed SE2E-based model.

\paragraph{Results Analysis}

\begin{figure}[ht]
  \centering
  \includegraphics[scale=0.85]{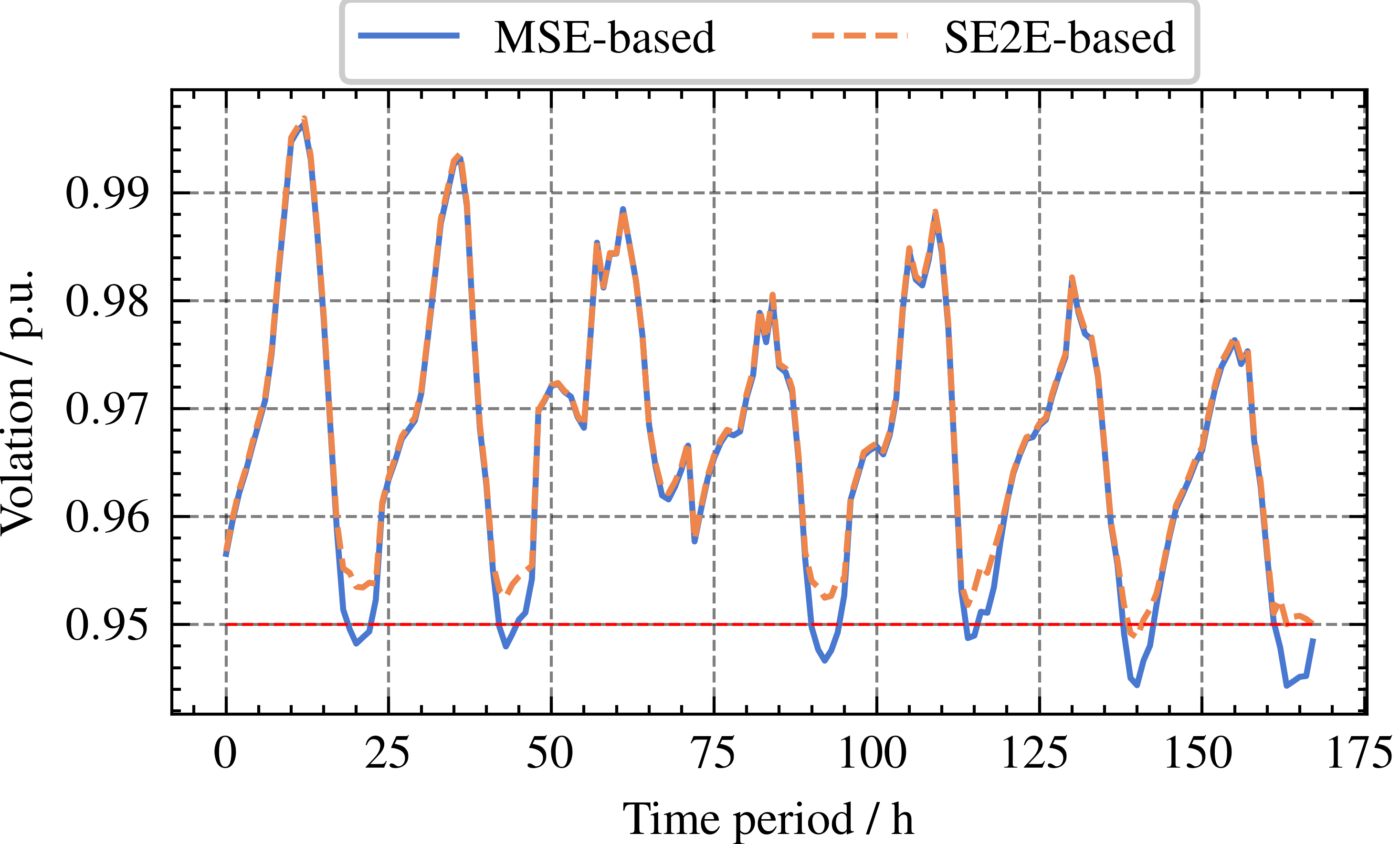}
  \caption{Bus voltage of bus 15 in a typical week.}
  \label{fig: volt}
\end{figure}

Fig. \ref{fig: volt} compares the voltage change of bus 15 in a typical week of the SE2E-based and MSE-based models. {\rv During the week, the number of voltage violations under the MSE-based model is higher than under the SE2E-based model. The proposed SE2E-based exhibits a more secure voltage operation than the MSE, verifying its safety awareness.} 

Tab. \ref{tab: pred safety} and \ref{tab: pl safety} further compare the prediction {\rv accuracy}, power loss, and the violation rate of the MSE-based and SE2E-based methods for coordinating prediction and decision evaluating when the penalty of voltage {\rv violation takes effect}.

Similar to the economic scenario analysis, Tab. \ref{tab: pred safety} compares the MAPE of multiple prediction models under the above two models in six test cases for prediction evaluation. As highlighted in {\rv Tab. \ref{tab: pred safety}}, the {\rv SE2E-based} shows higher prediction accuracy in PV4 and PV5 with lower MAPE. The overall prediction accuracy under {\rv MSE-based} is higher than under {\rv SE2E-based} with a lower MAPE. {\rv It demonstrates that due to the enforcement of the violation penalty, the shrinkage of voltage range degrades the overall multiple individual prediction accuracy.}

\begin{table*}[ht]
  \renewcommand{\arraystretch}{1.3}
  \centering
  \caption{Prediction performance analysis in different datasets in the safety scenario.}
  \begin{tabular}{ccccccccccccccc}
    \hline
    \multicolumn{1}{c}{\multirow{2}{*}{Case}} & \multicolumn{2}{c}{MAPE of PV1} & \multicolumn{2}{c}{MAPE of  PV2} & \multicolumn{2}{c}{MAPE of  PV3} & \multicolumn{2}{c}{MAPE of PV4} & \multicolumn{2}{c}{MAPE of PV5} & \multicolumn{2}{c}{MAPE of  PV6} & \multicolumn{2}{c}{MAPE of Load} \\ \cline{2-15} 
    \multicolumn{1}{c}{} & MSE & SE2E & MSE & SE2E & MSE & SE2E & MSE & SE2E & MSE & SE2E & MSE & SE2E & MSE & SE2E \\ \hline
    1 & 0.1890 & 0.2181 & 0.2997 & 0.4085 & 0.2881 & 0.2857 & 0.3075 & \textbf{\notice 0.3052} & 0.2948 & \textbf{\notice 0.2942} & 0.1927 & 0.2007 & 0.0490 & 0.0496 \\ 
    2 & 0.1941 & 0.2131 & 0.2686 & 0.3558 & 0.2578 & 0.2602 & 0.3171 & \textbf{\notice 0.3136} & 0.2973 & \textbf{\notice 0.2964} & 0.1985 & 0.2170 & 0.0451 & 0.0459 \\ 
    3 & 0.1930 & 0.2160 & 0.3005 & 0.4020 & 0.2966 & 0.3123 & 0.3091 & \textbf{\notice 0.3025} & 0.2846 & \textbf{\notice 0.2837} & 0.2103 & 0.2269 & 0.0477 & 0.0470 \\ 
    4 & 0.2243 & 0.2564 & 0.3342 & 0.4332 & 0.2844 & 0.3024 & 0.3160 & \textbf{\notice 0.3169} & 0.2642 & \textbf{\notice 0.2626} & 0.1980 & 0.2207 & 0.0470 & 0.0515 \\ 
    5 & 0.2221 & 0.2221 & 0.3072 & 0.4127 & 0.2786 & 0.2766 & 0.3116 & \textbf{\notice 0.3090} & 0.3008 & \textbf{\notice 0.2996} & 0.2265 & 0.2457 & 0.0583 & 0.0610 \\ 
    6 & 0.1767 & 0.1954 & 0.2815 & 0.3873 & 0.2538 & 0.2441 & 0.2771 & \textbf{\notice 0.2752} & 0.3166 & \textbf{\notice 0.3158} & 0.2157 & 0.2471 & 0.0484 & 0.0471 \\ \hline
    Avg. & 0.1999 & 0.2202 & 0.2986 & 0.3999 & 0.2765 & 0.2802 & 0.3064 & \textbf{\notice 0.3037} & 0.2930 & \textbf{\notice 0.2921} & 0.2069 & 0.2263 & 0.0492 & 0.0503 \\ \hline
    \end{tabular}
  \label{tab: pred safety}
\end{table*}

Tab. \ref{tab: pl safety} compares the power loss and violation of the above two models in the same six test cases for decision quality evaluation. The {\rv SE2E-based} shows an average 0.04kW higher power loss but an average 5.12\% lower violation rate than the {\rv MSE-based}, verifying the decision-making safety improvement {\rv from multiple predictions}. {\rv The SE2E-based sacrifices {\rv some economy of power loss and prediction accuracy} to achieve more safe predictions for more safe regulation.}

\begin{table}[ht]
  \renewcommand{\arraystretch}{1.3}
  \centering
  \caption{Decision performance analysis of different cases in the safety scenario.}
  \begin{tabular}{cccccc}
    \hline
    \multirow{2}{*}{Case} & \multicolumn{3}{c}{Power loss/kW} & \multicolumn{2}{c}{Violation rate/\% } \\ \cline{2-6} 
     & MSE & SE2E & Oracle & MSE & SE2E \\ \hline
     1 & 30.35 & 30.40 & 30.25 & 6.54 & 2.08 \\
     2 & 30.13 & 30.17 & 30.03 & 6.38 & 2.50 \\
     3 & 32.68 & 32.68 & 32.58 & 8.12 & 3.29 \\
     4 & 30.22 & 30.28 & 30.14 & 9.97 & 4.39 \\
     5 & 28.95 & 29.00 & 28.83 & 8.43 & 3.40 \\
     6 & 32.26 & 32.29 & 32.16 & 11.26 & 4.30 \\ \hline
     Avg. & 30.76 & 30.80 & 30.67 & 8.45 & 3.33 \\ \hline
    \end{tabular}
  \label{tab: pl safety}
\end{table}

Case study on \textit{safety scenario} that the prediction accuracy does not coincide with the decision quality from the safety aware perspective. The proposed SE2E endures higher PV prediction errors in multiple PV and load predictions. However, it improves decision-making safety with a lower violation rate, verifying the necessity of considering the downstream decision safety.

{\rv
\subsubsection{Discussion}

To summarize, the proposed SE2E-based model focuses on higher decision quality with lower power loss in Tab. \ref{tab: pl economic} and lower violation rate in Tab. \ref{tab: pl safety}, which may compromise the individual prediction accuracy but can enhance multiple prediction models' decision quality coordinately. Different parts of the \textit{regulation loss} in \eqref{eq: reg loss} take the main effect in different scenarios. So we verify the effectiveness of the proposed model in the economic and safety scenarios. In the economic scenario, the voltage violation penalty does not take effect, and the proposed semi-end-to-end model focuses on minimizing the power loss and achieves lower power loss than the conventional prediction, as shown in Tab. \ref{tab: pl economic}; in the safety scenario, the voltage violation penalty takes effect, and the proposed model focuses on reducing the voltage violation rate and achieves a lower violation rate, though with higher power loss than the conventional prediction, as shown in Tab.~\ref{tab: pl safety}.
}

{\rv
\subsection{Scalability and Compatibility Analysis}

The proposed semi-end-to-end model is scalable for various regulation devices and compatible with other objectives when the voltage regulation models are SOCP-driven models. The voltage regulation decision in the SOCP-driven voltage regulation model can come from the various regulation devices, including the SVC and inverter-based flexible resources, which can represent many renewable energy resources. So we add the $\hat{q}^{dg}_{j,t}$ from the six inverter-based PVs to the constraints for regulation scalability and the voltage regulation cost/inverter loss to the objective for multi-objective compatibility. The detailed decision-making and decision-evaluating of the scalable and compatible model is attached in the appendix \ref{subsec: mobj vcc}.

Tab. \ref{tab: pl mobj} compares the power loss, violation rate, voltage deviation, and regulation capacity of the SE2E-based with the MSE-based and oracle under the \eqref{eq: mobj dm}.
\begin{table}[ht]
  \renewcommand{\arraystretch}{1.3}
  \centering
  \caption{\rv Decision performance analysis in the multiple regulation and objectives scenario.}
  \begin{tabular}{cccc}
    \hline
    & {\rv MSE} & {\rv SE2E} & {\rv Oracle} \\
    \hline
    {\rv Power loss/kW} & {\rv 30.66} & {\rv 30.68} & {\rv 30.65}  \\
    {\rv Violation rate/\%} & {\rv 6.24}  & {\rv 2.86} & {\rv 0.00}  \\
    {\rv Voltage deviation/\%} & {\rv 0.3902}  & {\rv 0.3783}  & {\rv0.3784}  \\
    {\rv Regulation capacity/p.u.} & {\rv 25.55} & {\rv 26.07}  & {\rv 25.85} \\
    \hline
  \end{tabular}
  \label{tab: pl mobj}
\end{table}
Similar to the safety scenario, the proposed SE2E-based achieves a lower violation rate with higher power loss. Though the regulation capacity of the SE2E-based is higher than the others, the consideration of voltage deviation penalty in the objective function exhibits a 0.3783\% voltage deviation, lower than the MSE-based (0.3902\%) and oracle (0.3784\%). It verifies the safety awareness of the proposed model with the multiple regulation devices and multiple objectives set.

The above study verifies the regulation scalability and objective compatibility of the proposed semi-end-to-end model.
}

\section{Conclusion}\label{sec: conclusion}

This paper proposes a safety-aware semi-end-to-end coordinated decision model for voltage regulation by connecting the NN-driven prediction and SOCP-driven regulation models for high decision quality. \textit{Regulation loss} is designed to evaluate the economy and safety awareness of decision-making. Then the HSGD learning algorithm is proposed to train the multiple PV/load prediction models in a coordinated differential way. Case studies verify that compared to the conventional predict-and-optimize model, the proposed decision model achieves the regulation economy with lower power loss in the economic scenario and the safety awareness with a lower voltage violation rate in the safety-aware scenario. {\rv Future work may include the decision quality enhancing under state estimation uncertainty.}

{\rv
\appendix

\subsection{Prediction Model} \label{subsec: nn struct}

\subsubsection{Stochastic Gradient Descent Training}
Conventionally, the objective of each prediction model is to minimize the prediction error between the predicted output and the ground truth output, measured by MSE numerically in \eqref{eq: mse loss}. And the derivative of the MSE with respect to $\theta$ is calculated by \eqref{eq: mse derivative}.

\begin{equation}
    \mathcal{L}^{mse}_i = \frac{1}{|\mathcal{T}|} \sum_{t \in \mathcal{T}} (\hat{p}^{pred}_{i,t} - p^{pred}_{i,t})^2.
  \label{eq: mse loss}
\end{equation}

\begin{equation}
    \mathsf{D}_{\theta_i} \mathcal{L}^{mse}_i = 2  \sum_{t \in \mathcal{T}} (\hat{p}^{pred}_{i,t} - p^{pred}_{i,t}) \mathsf{D}_{\theta_i} \hat{p}^{pred}_{i}.
  \label{eq: mse derivative}
\end{equation}

\subsubsection{Neural Network Structure}

The NN is utilized to formulate the prediction models due to its prominent representational capacity. \eqref{eq: nn struct} illustrates the NN structure of $K$-layer full connected neural networks, mainly composed of the stacked linear transformation and activation functions. 

\begin{subequations}
  \begin{align}
    & x^k = \sigma(W_{k} x^{k-1} + b_k) \quad k = 1, ..., K\\
    & x^0 = x_i, \quad \hat{p}^{PV}_i = x^K
  \end{align}
  \label{eq: nn struct}
\end{subequations}
where $W_k$, $b_k$ are the linear transformation, bias of the $k$-th layer; $\sigma(\cdot)$ is the activation function, such as sigmoid, ReLU, and LeakyReLU.

\subsection{Voltage Regulation Model} \label{subsec: vcc}
The detailed explanation of the SOCP-driven voltage regulation models is presented as follows:
\begin{subequations}
  \small
  \allowdisplaybreaks
  \begin{align}
    \min_{q^{reg}} \quad & \qquad \sum_{t\in \mathcal{T}} \sum_{ij \in \mathcal{E}} l_{ij,t} r_{ij}  \label{socp a ap}\\
    \text{s.t.} \quad &\sum_{jk \in \mathcal{E}} P_{jk,t} - \sum_{ij \in \mathcal{E}} (P_{ij, t} - r_{ij}l_{ij,t}) = \hat{p}^{PV}_{j,t} - \hat{p}^{D}_{j,t}  \label{socp b ap}\\
    & \sum_{jk \in \mathcal{E}} Q_{jk,t} - \sum_{ij \in \mathcal{E}} (Q_{ij, t} - x_{ij}l_{ij,t}) = \hat{q}^{reg}_{j,t} - \hat{q}^{D}_{j,t}   \label{socp c ap}\\
    & v_{j,t} = v_{i,t} + (r^2_{ij}+x^2_{ij})l_{ij,t} - 2 (r_{ij}P_{ij,t} + x_{ij}Q_{ij,t}) \forall ij \in \mathcal{E}  \label{socp d ap}\\
    & || 2P_{ij,t} \quad 2Q_{ij,t} \quad l_{ij,t}-v_{i,t} || \leq l_{ij,t} + v_{i,t}  \label{socp e ap}\\
    & (V^{min}_{i,t})^2 \leq v_{i,t} \leq (V^{max}_{i,t})^2, \forall i \in \mathcal{N}_B  \label{socp f ap}\\
    & (I^{min}_{ij,t})^2 \leq l_{ij,t} \leq (I^{max}_{ij,t})^2, \forall ij \in \mathcal{E} \label{socp g ap} \\
    & q^{reg, min}_{i,t} \leq q^{reg}_{i,t} \leq q^{reg, max}_{i,t}, \forall i \in \mathcal{N}_{reg} \label{socp h ap}
  \end{align}
  \label{eq: voltage regulation opt ap}
\end{subequations}

The objective function \eqref{socp a ap} is to minimize the total power loss over the dispatch period $\mathcal{T}$. \eqref{socp b ap}-\eqref{socp e ap} formulates SOCP-driven distribution power flow model based on the second-order cone programming \cite{Li2020}. \eqref{socp b ap} and \eqref{socp c ap} are the bus power/var balance constraints, where $\mathcal{E}$ is the set of branches and $ij$ is the index of branch flowing from $i$ to $j$. \eqref{socp e ap} relaxes the nonlinear power flow equations by the second-order cone constraints. \eqref{socp f ap}-\eqref{socp h ap} limit the ranges of the voltage magnitude, transmission capacity, and var regulation output, where $\mathcal{N}_{B}$ and $\mathcal{N}_{reg}$ are the sets of buses and var regulators.

\subsection{Scalable and Compatible Voltage Regulation Model} \label{subsec: mobj vcc}

\subsubsection{Decision-making Layer of Scalable and Compatible Voltage Regulation Model}
We verify the regulation scalability and objective compatibility of the proposed semi-end-to-end model by adding $\hat{q}^{dg}_{j,t}$ from the inverter-based PV to the constraints and the voltage regulation cost/inverter loss to the objective, as follows:
\begin{subequations}
  \small
  \allowdisplaybreaks
  \begin{align}
    \min_{q^{reg}} & \sum_{t\in \mathcal{T}} \sum_{ij \in \mathcal{E}} l_{ij,t} r_{ij} {\rv + \alpha_1 \sum_{t\in \mathcal{T}} \sum_{i\in \mathcal{N}_{B}} (v_{i,t} - \hat{v_i})^2} \label{socp c.14 ra}\\
    & {\rv + \alpha_2 (\sum_{t\in \mathcal{T}} \sum_{i\in \mathcal{N}_{reg}} \hat{q}^{reg}_{j,t} + \sum_{t\in \mathcal{T}} \sum_{i\in \mathcal{N}_{dg}} \hat{q}^{dg}_{j,t})} \notag \\
    \text{s.t.} &\sum_{jk \in \mathcal{E}} P_{jk,t} - \sum_{ij \in \mathcal{E}} (P_{ij, t} - r_{ij}l_{ij,t}) = \hat{p}^{PV}_{j,t} - \hat{p}^{D}_{j,t} \label{socp c.14 rb}\\
    & \sum_{jk \in \mathcal{E}} Q_{jk,t} - \sum_{ij \in \mathcal{E}} (Q_{ij, t} - x_{ij}l_{ij,t}) = \hat{q}^{reg}_{j,t} + {\rv \hat{q}^{dg}_{j,t}} - \hat{q}^{D}_{j,t} \label{socp c.14 rc}\\
    & v_{j,t} = v_{i,t} + (r^2_{ij}+x^2_{ij})l_{ij,t} - 2 (r_{ij}P_{ij,t} + x_{ij}Q_{ij,t}) \forall ij \in \mathcal{E} \label{socp c.14 rd}\\
    & || 2P_{ij,t} \quad 2Q_{ij,t} \quad l_{ij,t}-v_{i,t} || \leq l_{ij,t} + v_{i,t} \label{socp c.14 re}\\
    & (V^{min}_{i,t})^2 \leq v_{i,t} \leq (V^{max}_{i,t})^2, \forall i \in \mathcal{N}_B\label{socp c.14 rf}\\
    & (I^{min}_{ij,t})^2 \leq l_{ij,t} \leq (I^{max}_{ij,t})^2, \forall ij \in \mathcal{E} \label{socp c.14 rg} \\
    & q^{reg, min}_{i,t} \leq q^{reg}_{i,t} \leq q^{reg, max}_{i,t}, \forall i \in \mathcal{N}_{reg} \label{socp c.14 rh} \\
    & {\rv \hat{q}^{dg}_{i,t} \leq \sqrt{S^{2}_{dg,i} - (\hat{p}^{PV}_{j,t})^2}, \forall i \in \mathcal{N}_{dg} } \label{socp c.14 ri}
  \end{align}
  \label{eq: mobj dm}
\end{subequations}
where $\alpha_1$ and $\alpha_2$ indicates the penalty coefficients of the voltage regulation cost and inverter loss; $\hat{q}^{dg}_{j,t}$ is the inverter-based distributed generator (DG) var regulation; $S^{2}_{dg,i}$ is the DG capacity.

\subsubsection{Decision-evaluating Layer of Scalable and Compatible Voltage Regulation Model}

The corresponding decision-evaluating layer of \eqref{eq: mobj dm} is formulated as follows:
\begin{subequations}
  \small
  \allowdisplaybreaks
  \begin{align}
    \min_{\mathcal{L}^{loss}} & \mathcal{L}^{loss} + \lambda \mathcal{L}^{penalty} \\
    \text{s.t. } & \mathcal{L}^{loss} \geq \sum_{t\in T} \sum_{ij \in E} l_{ij,t} r_{ij} + \alpha_1 \sum_{t\in \mathcal{T}} \sum_{i\in \mathcal{N}_{B}} (v_{i,t} - \hat{v_i})^2 \\
    & + \alpha_2 (\sum_{t\in \mathcal{T}} \sum_{i\in \mathcal{N}_{reg}} \hat{q}^{reg}_{j,t} + \sum_{t\in \mathcal{T}} \sum_{i\in \mathcal{N}_{dg}} \hat{q}^{dg}_{j,t}) \notag \\
    & \mathcal{L}^{penalty} = \text{ReLU}(v-(V^{max}_{i,t})^2) + \text{ReLU}((V^{min}_{i,t})^2-v)  \\
    & \sum_{jk \in E} P_{jk,t} - \sum_{ij \in E} (P_{ij, t} - r_{ij}l_{ij,t}) = p^{PV}_{j,t} - p^{D}_{j,t} \\
    & \sum_{jk \in E} Q_{jk,t} - \sum_{ij \in E} (Q_{ij, t} - x_{ij}l_{ij,t}) = \hat{q}^{reg}_{j,t} + {\rv\hat{q}^{dg}_{j,t}} - q^{D}_{j,t} \\
    & v_{j,t} = v_{i,t} + (r^2_{ij}+x^2_{ij})l_{ij,t} - 2 (r_{ij}P_{ij,t} + x_{ij}Q_{ij,t}) \\
    & || 2P_{ij,t} \quad 2Q_{ij,t} \quad l_{ij,t}-v_{i,t} || \leq l_{ij,t} + v_{i,t} 
  \end{align}
  \label{eq: mobj de}
\end{subequations}

}

\ifCLASSOPTIONcaptionsoff
  \newpage
\fi

\newpage

\begin{IEEEbiography}[{\includegraphics[width=1in,height=1.25in,clip,keepaspectratio]{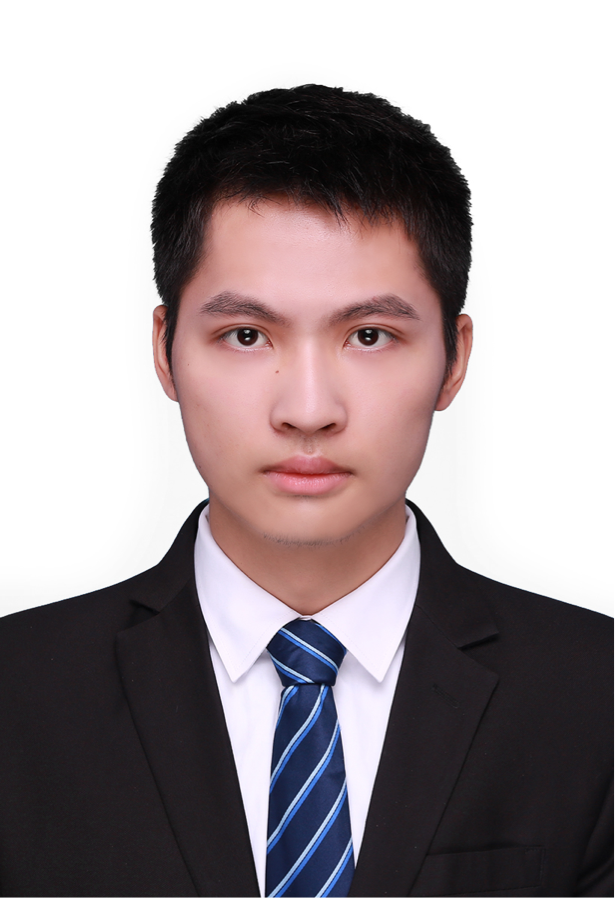}}]{Linwei Sang}(S'20) 
  received the M.S. degree from the School of Electric Engineering, Southeast University, China in 2021. 
  
  He is currently pursuing the Ph.D. degree in the Tsinghua-Berkeley Shenzhen Institute, Tsinghua University, Shenzhen, China. His research includes machine learning application in smart grid, the control of the distributed energy, and demand side resource management. 
\end{IEEEbiography}

\begin{IEEEbiography}[{\includegraphics[width=1in,height=1.25in,clip,keepaspectratio]{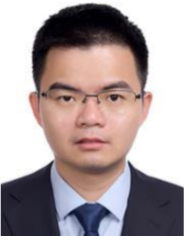}}]{Yinliang Xu}(SM'19)
  received the B.S. and M.S. degrees in control science and engineering from the Harbin Institute of Technology, Harbin, China, in 2007 and 2009, respectively, and the Ph.D. degree in electrical and computer engineering from New Mexico State University, Las Cruces, NM, USA, in 2013.
  
  He is currently an Associate Professor with Tsinghua-Berkeley Shenzhen Institute, Tsinghua Shenzhen International Graduate School, Tsinghua University, Beijing, China. His research interests include distributed control and optimization of power systems, renewable energy integration, and microgrid modeling and control. 
\end{IEEEbiography}

\begin{IEEEbiographynophoto}{Huan Long}(M'15)
  received the B.Eng. degree from Huazhong University of Science and Technology, Wuhan, China, in 2013, and the Ph.D. degree from the City University of Hong Kong, Hong Kong, in 2017. 
  
  She is currently an Associate Professor with the School of Electrical Engineering, Southeast University, Nanjing, China. Her research fields include artificial intelligence applied in modeling, optimizing, monitoring the renewable energy system and power system. 
\end{IEEEbiographynophoto}

\begin{IEEEbiographynophoto}{Wenchuan Wu}(Fellow, IEEE) received the B.S., M.S., and Ph.D. degrees from the Electrical Engineering Department, Tsinghua University, Beijing, China, where he is currently a Professor and the Deputy Director of the State Key Laboratory of Power Systems. His research interests include Energy Management System, active distribution system operation and control, machine learning and its application in energy system. He was a recipient of the National Science Fund of China Distinguished Young Scholar Award in 2017.
\end{IEEEbiographynophoto}

\end{document}